\documentclass[12pt]{iopart}
\usepackage{iopams}
\usepackage{bm}
\usepackage{amsfonts}
\usepackage{amsthm}
\usepackage{mathtools}
\usepackage{enumerate} 
\usepackage{graphicx}
\usepackage{dcolumn}
\usepackage[utf8]{inputenc}
\usepackage{xcolor}
\usepackage{xspace} 
\usepackage{soul} 
\usepackage[normalem]{ulem} 
\usepackage{hyperref}
\hypersetup{
    colorlinks=true,
    linkcolor=blue,    
    urlcolor=cyan,
    citecolor=blue,
}
\usepackage{subfigure}
\usepackage{subcaption}
\usepackage{caption}
\usepackage{color}
\date{\small \today}

\begin{document}
\newcommand{\ZSQ}[1]{{\color{red}#1}}
\newcommand{\figref}[1]{{Figure \ref{#1}}}

\title{Extracting a stochastic model for predator-prey dynamic of turbulence and zonal flows with limited data}

\author{J. C. Huang\textsuperscript{1}, Z. S. Qu\textsuperscript{1,*}, R. Varennes\textsuperscript{1}, Y. W. Cho\textsuperscript{1}, X. Garbet\textsuperscript{1,2}, C. G. Wan\textsuperscript{1}, C. Guet\textsuperscript{1}, D. Niyato\textsuperscript{3}, V. Grandgirard\textsuperscript{2}
}

\address{$^1$School of Physical and Mathematical Sciences, Nanyang Technological University,  Singapore 637371, Singapore}
\address{$^2$CEA, IRFM, F-13108 Saint-Paul-Lez-Durance, France}
\address{$^3$College of Computing and Data Science, Nanyang Technological University, Singapore 639798, Singapore}
\ead{zhisong.qu@ntu.edu.sg}
\vspace{10pt}

\begin{abstract}
Understanding the interaction between turbulence and zonal flows is critical for modeling turbulence transport in fusion plasmas, often described through predator-prey dynamics. However, traditional deterministic models like the Lotka–Volterra equations simplify this interaction and fail to capture the small fluctuations in simulation data. In this study, we develop a neural network model based on stochastic differential equations (SDEs) to represent the predator-prey dynamics using limited data from simulations of the modified Hasegawa–Wakatani system. We extract the drift and diffusion terms via neural networks, incorporating physical constraints and employing the unscented transform to mitigate challenges brought by limited data. The model accurately reproduces key dynamical features, including stagnation phenomena and energy exchange mechanisms, and the state density distribution generated from the model shows a low KL divergence with the simulation data. A parameter scan reveals that zonal flow shearing efficiency decreases with amplitude, and predator-prey oscillations damp in the absence of stochasticity. These findings underscore the value of integrating physical insight into data-driven approaches for complex plasma systems.
\end{abstract}
\section{Introduction}

    \indent Dynamical systems in plasma physics, such as dynamics in Tokamaks, are inherently complex. This complexity is further amplified by their high degree of nonlinearity. Reduced models are then crucial for understanding the mechanism that govern their dynamics, and they have already proven to be effective in fields where numerical solutions are difficult to obtain and expensive, due to the non-linear effects
    ~\cite{zhang_advances_2018,li_surrogate_2024,oleary_stochastic_2022}. Deep learning is a promising tool for assisting in the development of reduced models, even when the underlying physics is highly complex. Recent advances in applying deep learning to reduced modeling in plasma physics have shown considerable potential. In particular, the use of generative models such as Generative Adversarial Networks (GANs) \cite{GAN} has enabled the construction of accurate and interpretable surrogates for turbulent fields, capturing essential spatial and statistical features of plasma dynamics \cite{clavier_generative_2024, castagna_stylegan_2024, gahr_scientific_2024}. These developments build upon earlier work that applied neural networks to reduced models of Hasegawa–Wakatani turbulence, demonstrating the ability of data-driven methods to extract coherent structures and learn the underlying dynamics from simulation data \cite{heinonen_learning_2020, heinonen_turbulence_2020}.
    In this paper, we focus on a reduced model for the dynamic between turbulence and zonal flow, as their interplay governs energy transport and self-regulation mechanisms in magnetized plasmas, directly impacting confinement performance in fusion devices. Building on early gyro-kinetic studies and bifurcation analyses, which emphasized the critical role of coherent structures in shaping confinement transitions \cite{lin_turbulent_1998, chen_excitation_nodate, yoshida_rapid_2003}, theoretical models inspired by predator-prey dynamics have provided valuable insight into the self-regulation mechanisms between turbulence and zonal flows, revealing how energy is exchanged and organized in magnetized plasmas \cite{diamond_self-regulating_1994, smolyakov_zonal_2000, kim_dynamics_2002, chen_zonal-flow_2004, miki_novel_2011}.\\
    \indent The predator-prey model of turbulence and zonal flow is a simple
    but efficient model, given by the Lotka–Volterra equations
    \begin{align}\label{PP}
        &\frac{dE}{dt} = \alpha E - \beta EU, \nonumber\\
        &\frac{dU}{dt} = -\gamma U + \sigma EU,
    \end{align}
    where \(E\) and \(U\) are the energies associated with turbulence and zonal flow shear, respectively. The coefficients \(\alpha\), \(\beta\), \(\gamma\) and \(\sigma\) are positive constants. 
    Here, $E$ grows exponentially with growth rate $\alpha$ due to linear drift-wave instability, representing the ``prey''.
    The zonal flow shearing effect suppresses the turbulence via the cross term $-\beta E U$, representing the ``predator''. 
    Meanwhile,  the zonal flows are driven via Reynolds stress $\sigma EU$, and linearly damped due to collisional effects $-\gamma U$.
    This model and its variant were successfully applied to capture the dynamic between the turbulence energy and the zonal flow shear energy~\cite{diamond_zonal_2005,itoh_physics_2006,itoh_physics_2008,itoh_coherent_2005}.

    Despite the qualitative success of the predator-prey model in describing turubencle-zonal flow interaction, a question remains: can we directly extract \eqref{PP} from the simulation data of turbulence and zonal flows as well as the corresponding parameters without knowing its form?
    An example is given in \figref{SDEs1.png} showing the time evolution of turbulence energy and zonal flow energy. These data are extracted from a direct simulation of the modified Hasegawa-Wakatani equations detailed in Section \ref{sec:mHW}.
    Even though the signature of Lotka–Volterra is clear as the zonal flow energy lags behind the turbulence energy, the amplitude and period of oscillation changes from peak to peak.
    The reason behind this is that the dynamics between turbulence and zonal flow is complicate, as turbulence itself is inherently chaotic and nonlinear. 
    Some successful model extraction was reported in \cite{koba}, but in a scenario with limited randomness.
    Other challenges may also present.
    First, the coefficients may not be constant in time as the zonal flows evolve, especially when zonal flows merge into large and global ``staircases'' over time.
    Second, the Lotka–Volterra equations are over-simplified. It was shown that the coefficients $\beta$ and $\sigma$ are functions of $E$ and $U$ \cite{diamond_zonal_2005}.
    Meanwhile, other non-linear terms might present, for instance the non-linear turbulence self-interaction could give a term proportional to $E^2$ in the first equation \cite{diamond_zonal_2005}.
    Finally, even if one is able to extract a model by some mean, a quantitative validation of the extracted model is difficult, given that the small fluctuation prevent an exact reproduction of $E$ and $U$ versus time in data.
    
    To capture features in such a system by a physics model, one attempt is to generalize coefficients $\alpha, \beta, \gamma$ and $\sigma$ to functions of $E$ and $U$, and to introduce terms that captures small fluctuations in data, in our case stochastic terms. Indeed,
    stochastic predator-prey model \cite{tb_sde} is suitable for capturing the dynamics of turbulence and zonal flow~\cite{Kim0}, as it have succeeded in capture features of population dynamic \cite{dobramysl_stochastic_2018} and ordinary differential equation \cite{dam_sparse_2017}. 
    In this paper, we postulate the interaction between turbulence and zonal flow to take the form of stochastic differential equations (SDEs) given by
    \begin{align}\label{SDEs}
        &dE = g_{11}(E,U)dt + g_{21}(E,U)dw_1, \\
        \label{SDEs1}
        &dU = g_{12}(E,U)dt + g_{22}(E,U)dw_2,
    \end{align}
    where \(g_{11}\) and \(g_{12}\), are drift terms while \(g_{21}\) and \(g_{22}\) are terms related to stochasticity that captures small fluctuations. The terms \(dw_1\) and \(dw_2\) are Brownian motions (can be correlated or independent). The unknown functions \(g_{11}(E,U)\), \(g_{12}(E,U)\), \(g_{21}(E,U)\) and \(g_{22}(E,U)\) need to be learned from data.
    \\
    \indent Such a stochastic model is promising in extracting features from large amounts of data, but shows high error when the amount of data is small~ \cite{wang_data-driven_2022,oleary_stochastic_2022}. However, data generation from simulation is usually expensive and data scarcity would lead to inaccuracies. We address such limitation on data size by taking grid average and setting input data threshold to only use data with enough statistics. To make learned non-linear function robust, we also used unscented transform~\cite{julier_new_nodate,haykin_unscented_2001,leslie_properties_1960}, a technique that corrects the bias due to insufficient data. Neural network is a suitable tool for representing the unknown $g$ functions and has been shown to possess the potential to learn high-dimensional nonlinear effect~\cite{zhu_bayesian_2018}. Physics-informed neural network is a trending topic for surrogate model development, which adopt three loss terms to incorporate PDEs, initial conditions and boundary conditions for a specific system~\cite{raissi_physics-informed_2019}. In our study, physics-based design of the network and losses are also introduced to guide neural network to learn. Those introduced hints enforce known physics into stochastic model.
    Finally, we validate the extracted SDE by comparing the distribution of states of simulations using the SDE and real data, and quantifying it with the Kullback–Leibler divergence. The model results show a good agreement with the data, as they not only capture the trajectory "stagnation" feature but also shows a similar distribution, which will be explained later.
    
    This paper is organized as follows. Section 2 presents the theoretical background of our study, including the modified Hasegawa–Wakatani equations, stochastic differential equations, data pre-processing, the basic structure of the adopted neural network, and evaluation methods.
    Section 3 outlines the procedure used to improve the model's capacity to learn relevant features through some assistance, along with the corresponding results for one set of simulation data.
    Section 4 examines the dependency of the learned model on the density gradient - the drive of the system.
    Finally, Section 5 concludes the paper and discusses possible future directions.

\section{Theory and methods}
    In this section, we introduce how the data are generated and how the components \(g_{11}\), \(g_{12}\), \(g_{21}\), and \(g_{22}\) in equation \eqref{SDEs} and \eqref{SDEs1} are extracted from the data. We begin by presenting the modified Hasegawa–Wakatani model, followed by our stochastic differential equation (SDE) framework, where we explicitly describe the relationship between the model parameters and the data. Next, we explain how the drift and diffusion functions are represented using neural networks. Finally, we describe the methods used to evaluate the performance of our model.

\subsection{Raw Data preparation}
\label{sec:mHW}
     In this section, we introduce how the raw data for training is generated. The data is produced using TOKAM2D \cite{sarazin_intermittent_1998,phillip_2022}, a reduced 2D fluid simulation code designed to describe electrostatic turbulence in magnetized plasmas. TOKAM2D uses spectral methods for spatial dimensions, and fourth-order Runge-Kutta for time advection. It is GPU compatible – in our case a simulation takes 2 hours on a single A100 card on average. TOKAM2D captures essential features of drift-wave dynamics and the self-organization of zonal flows, making it a valuable tool for investigating transport processes and testing reduced modeling approaches in simplified geometries. In our case, TOKAM2D is based on the modified Hasegawa–Wakatani (mHW)~\cite{hasegawa_plasma_1983} equations, which are particularly suitable for studying predator–prey models, as they can describe the evolution of zonal modes.  
    
    The mHW equations are arguably the simplest paradigm to study the interaction between drift-wave turbulence and zonal flows. 
    The equations couple the perturbed density of the plasma, $n$, and the perturbed electrostatic potential, $\phi$. In 2D, the coordinates $x$ and $y$ are the radial coordinate and the poloidal direction coordinate, respectively, both with periodic boundary conditions. These equations are given by
    \begin{equation}\label{HWE}
        \begin{aligned}
            \frac{\partial \zeta}{\partial t} + \{\phi, \zeta\} &= \alpha (\tilde{\phi} - \tilde{n}) - D \nabla^4 \zeta - \nu \langle \zeta \rangle_y, \\
            \frac{\partial n}{\partial t} + \{\phi, n\} &= \alpha (\tilde{\phi} - \tilde{n}) - \kappa \frac{\partial \phi}{\partial y} - D \nabla^4 n,
        \end{aligned}
    \end{equation}
     where $\zeta = \nabla^2 \phi$ is the vorticity, $\nabla^2$ the two-dimensional Laplacian operator in 2D,
     and $\langle \ldots \rangle_y$ stands for averaging over the $y$ direction. The term $\{\phi, n\} = \frac{\partial \phi}{\partial x} \frac{\partial n}{\partial y} - \frac{\partial n}{\partial x} \frac{\partial \phi}{\partial y}$ is the standard Poisson bracket operator. The hyper-diffusion coefficient $D$ is introduced for numerical stability. The drive of the system is specified by $\kappa = -\frac{\partial \ln n_0}{\partial x}$, the density gradient scale-length, where $n_0$ is the plasma density of the reference equilibrium.
     The non-zonal terms $\tilde{\phi}$ and $\tilde{n}$ are given by
     \begin{equation}
         \tilde{\phi} = \phi - \langle \phi \rangle_y,
     \end{equation}
     
     \begin{equation}
         \tilde{n} = n - \langle n \rangle_y.
     \end{equation}
     where the zonal component $\langle . \rangle_y$ is the average along the poloidal direction. The adiabatic parameter $\alpha$ is proportional to the plasma conductivity. In the collisionless case $\alpha \to \infty$, the density follows the Boltzmann relation $n = \phi$, and Eq.\eqref{HWE} becomes the Hasegawa-Mima equation \cite{HWMM}. In the highly collisional case $\alpha \to 0$, the HW equations resemble the incompressible 2D Navier-Stokes equation.
     In addition, a zonal flow damping term $-\nu \langle \zeta \rangle_y$ is added to mimic the collisional damping of zonal flow, which is crucial in the predator-prey dynamics, as discussed later. 
    The density is normalized by $\frac{L}{\rho_0} n_0$, where $L$ is the size of the box, $\rho_0$ is a reference Larmor radius. The potential is normalized by $\frac{L}{
    \rho_0} \frac{e}{T_e}$, where $e$ is the electron charge, and $T_e$ is the electron temperature (supposed constant in this model). In this paper we take normalized value as default. 
    \begin{figure}[hbt!] 
        \centering
        \includegraphics[width=1.0\linewidth]{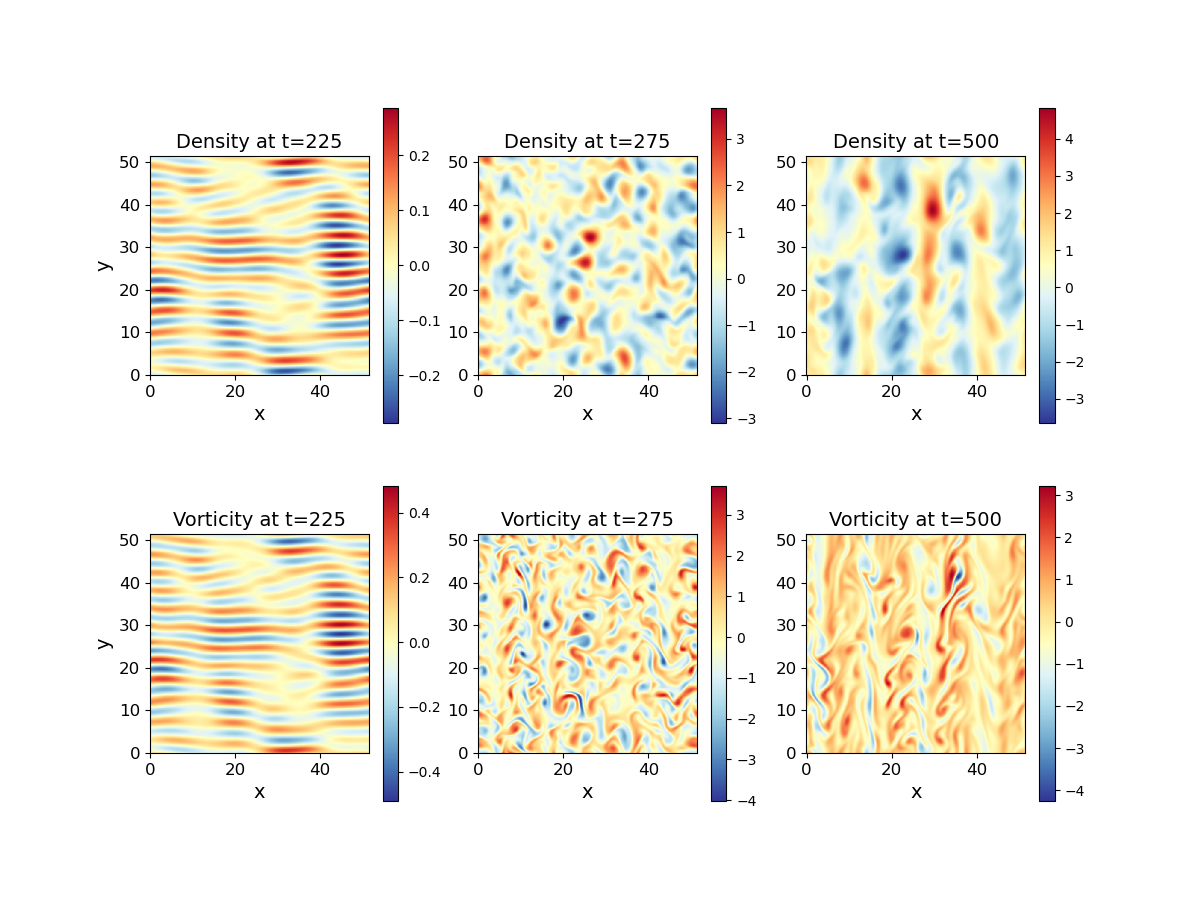} 
        \caption{The perturbed density $n$ and vorticity $\zeta$ from the simulation with mHW equations, at linear stage $t=225$ (left), early nonlinear stage $t=275$ (middle) and steady-state $t=500$ (right).(Here all value are normalized as previous mentioned)}
        \label{fig:simulation_result}
    \end{figure}
    \figref{fig:simulation_result} gives an example of the simulation results with $\kappa=1.05$, $\alpha=2$, $D=0.0001$ and $\nu=0.01$.
    The box size is $L_x=L_y=51.5{\rho_0}$, with a spatial resolution of $512\times 512$. The time step size is set to $2^{-9}\frac{L}{C_0}$ and the time interval between each output step $\Delta t= 0.25$.
    In the linear stage, the fluctuations are dominated by the fastest growing modes with a low wave-number in $x$, showing long stripes in the $x$ direction.
    When the system enters the nonlinear stage, zonal flows moving along the $y$ direction are generated, which tears the turbulent eddies apart into smaller eddies.
    Finally, in the nonlinear steady-state, strong and persistent zonal flows form, which separates the box into several vertical stripes, only allowing turbulence activities at the location where the zonal flow shear is nearly zero (maximum or minimum of zonal flow amplitude).

\begin{figure}[h] 
            \centering
            \includegraphics[width=0.6\textwidth]{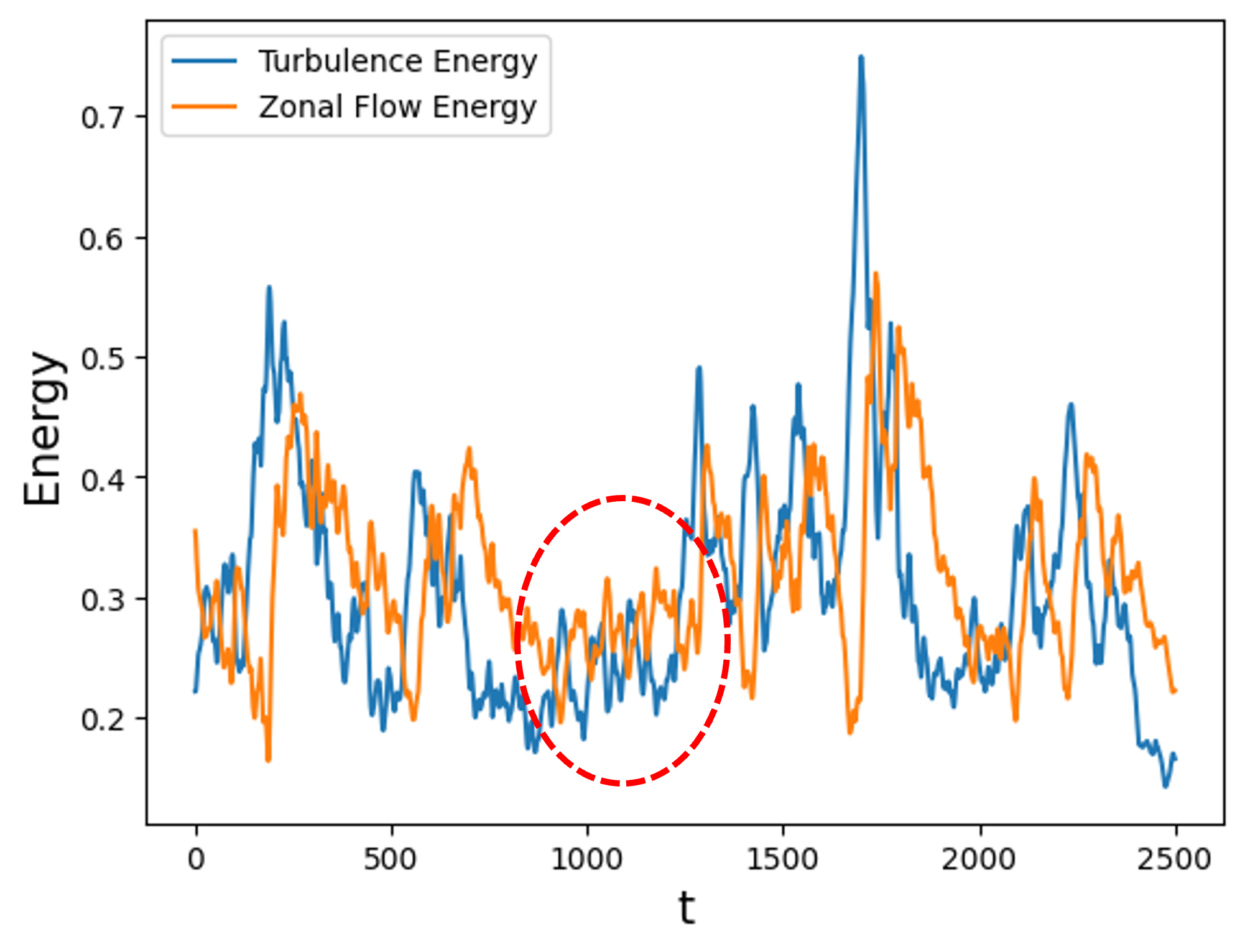} 
            \caption{Turbulence and zonal flow energy extracted from a simulation of the modified Hasegawa-Wakatani equations using the TOKAM2D code. A stagnation phenomenon is indicated by the red dash circle. All values are normalized, as detailed in Sec\ref{sec:mHW}}
            \label{SDEs1.png}
    \end{figure}

    In long time simulations, the zonal flow wave number changes over time, alternating between three and four peaks in the $x$ direction.
    To avoid the problem of changing zonal flow wave-number, hereafter we only simulate $1/4$ of the box in $x$ direction.
    We also only collect data after the zonal flow stair-cases are formed, discarding the data from the linear and early nonlinear stages, i.e. $t\le500$. 
    We then take the turbulence energy and the zonal flow energy as proxies for the predator and the prey, respectively. They are given by
    \begin{equation}
        E= \frac{1}{L_x L_y} \int \frac{1}{2} \left( |\nabla \tilde{\phi}|^2 + \tilde{n}^2 \right) \, dxdy,
        \tag{22}
    \end{equation}
    \begin{equation}
        U = \frac{1}{L_x} \int \frac{1}{2} \left| \langle \partial_x\phi \rangle_y \right|^2 \, dx,
        \tag{23}
    \end{equation}. 
    An example of data is given in \figref{SDEs1.png}. This data form a couple of trajectories of turbulence energy and zonal flow energy as a function of time.

\subsection{Stochastic differential equations (SDEs)}\label{sed}
    We adopt a stochastic predator–prey model to investigate the dynamics between turbulence and zonal flows. Compared to its deterministic counterpart, the stochastic formulation captures high-frequency features through the inclusion of a diffusion term that introduces randomness at each time step. Before further processing the raw data, we first introduce the framework of stochastic differential equations (SDEs). In our study, the general form of a SDE can be written as
    \begin{equation}\label{tadf}
        \begin{aligned}
            &d\mathbf{x} = \mathbf{g}_1(\mathbf{x}(t))dt + \mathbf{g}_2(\mathbf{x}(t))\odot d\mathbf{w}(t),\\
        \end{aligned}
    \end{equation}
    where \(\mathbf{x}(t) = (E,U)\) is the state of turbulence energy and zonal flow energy at time $t$, \(\mathbf{g}_1(\mathbf{x}(t))\) and \(\mathbf{g}_2(\mathbf{x}(t))\) are the drift and the random diffusion amplitude, respectively, \(d\mathbf{w}(t)\) the  vectorized Brownian motion, and $\odot$ the element-wise multiplication. 
    Hereafter, the vector \((E,U)\) is called a \textbf{state}, and the state as a function of $t$ is called a \textbf{trajectory}.
    
    In SDEs, \( d\mathbf{w} \) represents the infinitesimal increment of Brownian motion, which has zero mean and variance proportional to \( dt \). In numerical simulations, this is commonly approximated as \(d\mathbf{w}(t)\approx\mathbf{\nu}\sqrt{dt}\)~\cite{tb_sde} where \(\mathbf{\nu}\sim \mathcal{N}(0,\mathbf{\Sigma})\) is a normal distribution with zero mean and a covariance matrix of $\mathbf{\Sigma}$ (normalized to have ones on the diagonal) to be determined, giving the form 
    \begin{equation} \label{eq:gSDEs}
        \begin{aligned}
            &d\mathbf{x} \approx \mathbf{g}_1(\mathbf{x}(t))dt + \mathbf{g}_2(\mathbf{x}(t))\odot \mathbf{\nu}\sqrt{dt},\\
        \end{aligned}
    \end{equation}
    With a given initial condition, a trajectory \(\mathbf{x}(t)\) could be build. In this work, we use the Euler method to integrate a given SDE numerically, written as
        \begin{equation}\label{eq10}
            \mathbf{x}(t+\Delta t) \approx \mathbf{x}(t) + \mathbf{g_1}(\mathbf{x}(t))\Delta t + \mathbf{g_2}(\mathbf{x}(t)) \odot \mathbf{\nu} \sqrt{\Delta t}.
        \end{equation}
    where $\Delta t$ is a chosen time step.
    At each time, a random number $\nu$ is drawn, leading to different trajectories even with the same initial condition. 
        \\
    \indent Suppose the energy of turbulence and zonal flow in an mHW system follows an SDE given by Eq.\eqref{eq:gSDEs}, time drift function \(\mathbf{g}_{1}(\mathbf{x})\) could be extracted from data by taking repetitions of this simulation with the same initial state but different random seeds. To be specific, if one is allowed to choose the same initial state $\mathbf{x}(t)=(E,U)=\mathbf{\xi}$ at $t$, advancing such stochastic system from $t$ to $t +\Delta t$ gives $\mathbf{x}(t+\Delta t)$. If we assumes \(\Delta t\) to be small enough, with a very large number of repetitions by drawing the random variable $\nu$, one could estimate \(\mathbf{g}_{1}(\mathbf{x})\) by
    \begin{equation}\label{getg1}
        \mathbf{g}_1(\mathbf{\xi}) = \frac{1}{\Delta t}E[\mathbf{x}(t+\Delta t)-\mathbf{x}(t)|\mathbf{x}(t)=\mathbf{\xi}],
    \end{equation}where operator \(E\) represents conditional expectation over trajectories under condition that each trajectory starts at position \(\mathbf{\xi}\). With \(g_1(\mathbf{x})\) being computed, one could get random diffusion function \(g_{2}(\mathbf{x})\) by taking variance of increments over trajectories, given by 
    \begin{equation}\label{getg2}
            \mathbf{g}_2^2(\mathbf{\xi}) = \frac{1}{\Delta t}\text{Var}[\mathbf{x}(t+\Delta t)-\mathbf{x}(t)|\mathbf{x}(t)=\mathbf{\xi}],
    \end{equation}
    where $\mathbf{g}_2^2$ denotes the element-wise square. 
    Both expressions in Eq.\eqref{getg1} and Eq.\eqref{getg2} rely on the law of large numbers, as they require computing the mean and variance from data. To obtain accurate estimates of the drift and diffusion functions.

\subsection{Data processing}

    \indent In our case, for each set of physical parameters, we take 50 TOKAM2D simulations with different random initial perturbation in $n$ and $\phi$.
    As mentioned in Section \ref{sec:mHW}, the linear stage and early nonlinear stage are discarded.
    We then collect the data from $t=500$ to $t=3000$ with a time interval of output $\Delta t=0.25$, in total 10000 states per simulation.
    We note that this is still a small amount of data for the discovery of stochastic functions, as the amount of data for the accurate extraction of a stochastic model from such a complex chaos system would typically be about 10 times larger.\cite{oleary_stochastic_2022}.

    For these raw data, we need to perform further processing. 
    As mentioned previously, in theory the accuracy for \(\mathbf{g}_1(\mathbf{x})\) and \(\mathbf{g}_2(\mathbf{x})\) depends on the number of trajectories repeated from the same initial state of turbulence and zonal flow energy. However, it is generally not possible for simulation states from data to have a initial state exactly equal to $\mathbf{\xi}$. Therefore, to estimate \( \mathbf{g}_1 \) and \( \mathbf{g}_2 \) from trajectory data, we assume that we observe a large number of time series (trajectories) of \((E, U)\). By grouping trajectory segments based on similar past states \(\mathbf{x} = (E, U)\), we can compute the local average and variance of their increments. Specifically, \( \mathbf{g}_1(\mathbf{x}) \) is estimated as the conditional expectation of the time derivative of \(\mathbf{x}\) of similar states, and \( \mathbf{g}_2(\mathbf{x}) \) is estimated from the conditional variance.

    
    
    To be precise, we first discretize the entire 2D state domain into 40 equally sized square sections. All states falling within the same section are considered to share the same preceding state, which is represented by the center of the corresponding box.
    \begin{figure}[hbt!] 
        \raggedright
        \includegraphics[width=0.7\textwidth]{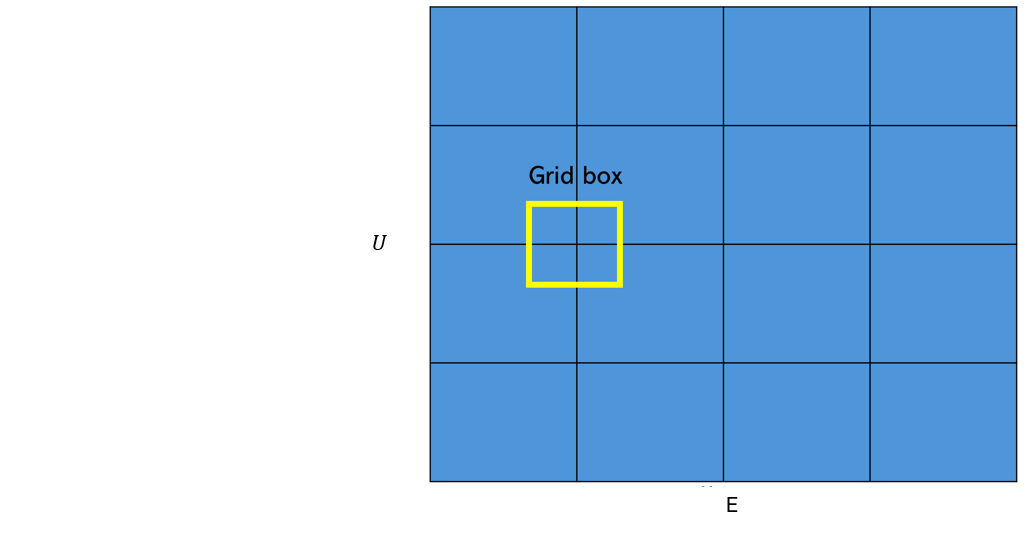} 
        \caption{Illustration of grid buckets: the states space are discretized into equally spaced grid points. The yellow box illustrates a grid buckets centered on one of these grid points, within which all data are assumed to originate from the same preceding state—the center of the grid box.}
        \label{grid.fig}
    \end{figure}
    As illustrated in \figref{grid.fig},  each grid point marks a state bucket that takes in near states observed. After sorting data in to buckets, each bucket contains on average \(10^3\) data points.
    \begin{figure}[hbt!] 
        \centering
        \includegraphics[width=0.8\textwidth]{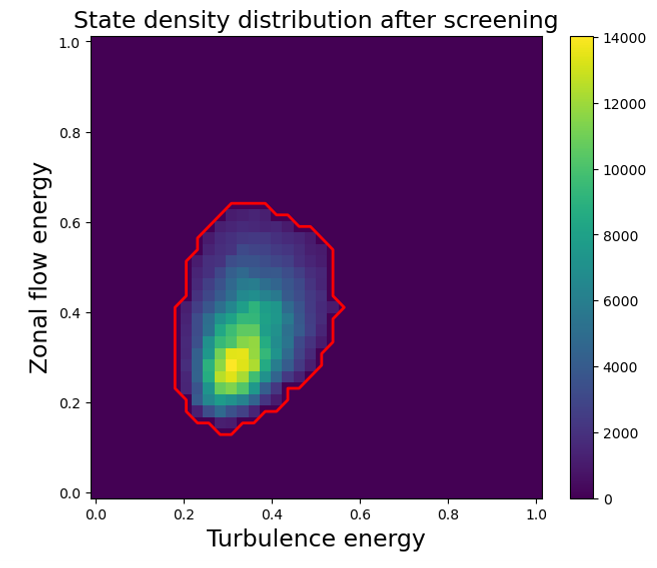} 
        \caption{The number of states in each grid box.
        The red contour shows the boundary where the number of data in the grid box exceeds 500.}
        \label{screen.fig}
    \end{figure}
    
    \figref{screen.fig} shows the distribution of states in each grid box.
    A close inspection of \figref{screen.fig} shows that the number of states in certain area of the 2D grid is great, while it is proven to be small or zero in other areas. 
    These edge states do not contain enough data to make statistical significance.
    We only make use of the grid sections (buckets) where the number of states exceeds a threshold of $500$, as this ensures sufficiently stable estimates of the expectation and variance. Below this threshold, the computed statistics become too unreliable due to insufficient sampling. 
    This is marked by the red doted line in \figref{screen.fig}.
    
    After sorting data into buckets, we can compute $\Delta \mathbf{x}(\mathbf{x}(t)) = \mathbf{x}(t+\Delta t) - \mathbf{x}(t)$ by taking the difference between the neighboring states in time. 
    We need to modify Eq.\eqref{getg1} and Eq.\eqref{getg2} since the assumption that the trajectories are starting from the same initial state is no longer true, as the simulated data cannot fully satisfy the same initial state conditions due to resolution limit.
    In fact, the data $\Delta \mathbf{x}$ contains both the randomness from $\mathbf{\nu}$, and the distribution of $\mathbf{x}$ within a box.
    Taking the expectation of $\Delta\mathbf{x}$ within a bucket, it gives
    \begin{equation}\label{ngetg1}
         E[\Delta\mathbf{x}, \text{for} \left| x_i - \xi_i \right|\le \rho,i=1,2] = E_\mathbf{x}[\mathbf{g}_1(\mathbf{x})] \Delta t,
    \end{equation}
    where $\mathbf{\xi}$ now is the value of $\mathbf{x}$ at the center of the bucket, $\rho$ is half of bucket size and the subscript ``$i$'' means the $i$-th element of the state vector \(\mathbf{x}(t)\).
    The expectation $E_\mathbf{x}$ averages over the distribution of $\mathbf{x}$ within a bucket.
    Taking the expectation within the bucket is equivalent to taking expectation of both $\mathbf{\nu}$ and $\mathbf{x}$.
    We note that the $\mathbf{g}_2$ part vanishes due to the independence between $\mathbf{x}$ and $\mathbf{\nu}$.
    The left-hand side of Eq.\eqref{ngetg1} can be estimated from data,
    while the right-hand side should be estimated from the modeled $\mathbf{g}_1(\mathbf{x})$, i.e. the neural network representation of $\mathbf{g}_1$.
    We use the unscented transform operator $\text{UT}$ \cite{UT,UT0,UT1} as a replacement of expectation over $\mathbf{g}_1$ to save computational cost, which converts the expectation into the weighted-average of five (in 2D) ``sigma-points'' (integration points) computed from the mean and variance of $\mathbf{x}$ within a bucket, given by
    \begin{equation}\label{14}
       E_\mathbf{x}[\mathbf{g}_1(\mathbf{x})] \approx \text{UT}[\mathbf{g}_1(\mathbf{x})] = \sum_{l=0}^{4} W_l \mathbf{g}_1(\mathbf{z}_l),
    \end{equation}
    where $W_i$ is the weight, and $\mathbf{z}_i$ are the sigma points within the buckets.
    The unscented transform is a Gauss-like quadrature mimicking the process of sampling.
    The detail can be found in \ref{app:UT}.
    
    Taking the variance of $\Delta \mathbf{x}$ within a bucket, one gets 
    \begin{equation}\label{ngetg2}
        \begin{aligned}
            \ \text{Var}[\Delta \mathbf{x},\text{for} &\left| x_i - \xi_i \right|\le \rho,i=1,2]\\
            &= E[(\Delta \mathbf{x} -  E_\mathbf{x}[\mathbf{g}_1(\mathbf{x})]\Delta t)^2,\text{for} \left| x_i - \xi_i \right|\le \rho,i=1,2] \\
            &= (E_\mathbf{x}[\mathbf{g}_1^2(\mathbf{x})] - E_\mathbf{x}[\mathbf{g}_1(\mathbf{x})]^2) \Delta t^2 + E_\mathbf{x}[\mathbf{g}_2^2(\mathbf{x}(t)]\Delta t.
        \end{aligned}
    \end{equation}
    The first term on the right-hand side of Eq.\eqref{ngetg2} is a correction due to non-zero box width.
    Again, the expectations over $\mathbf{x}$ are replaced by unscented transform, while the variance is computed from data.
   To avoid left-hand side variance becoming negative due to insufficient data,  we only keep the first term of the right-hand side if the number of samples is above a given threshold.
    
\subsection{Neural network (NN) structure and loss}
\label{sec:NN}
    \begin{figure}[hbt!] 
        \raggedright
        \includegraphics[width=1.\textwidth]{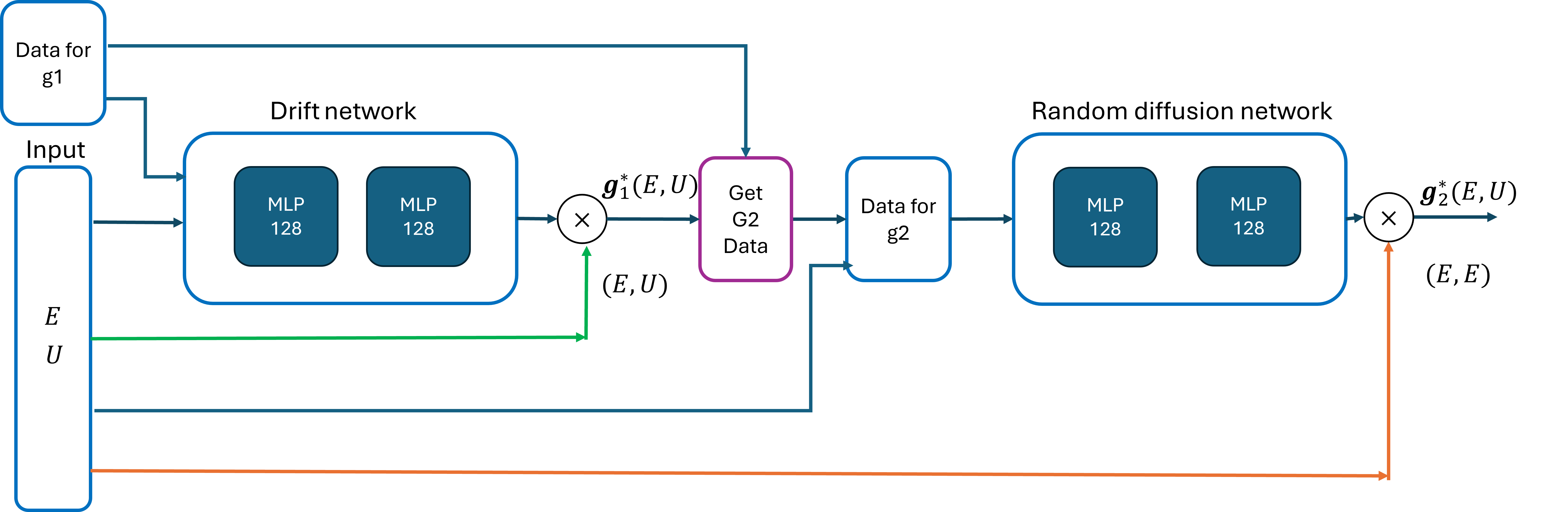} 
        \caption{NN structure, where \(E\) and \(U\) represent the energy of turbulence and the energy of zonal flow respectively. Orange and green lines represents direct connection to dot product operator, where \(\mathbf{g_1}^*(E,U)\) takes dot product with \((E,U)\) and \(\mathbf{g_2}^*(E,U)\) takes dot product with \((E,E)\).}
        \label{NN.fig}
    \end{figure}
    In the previous section, we obtained the right-hand sides of Eq.\eqref{14} and Eq.\eqref{ngetg2} from data. We now aim to represent the drift function \(\mathbf{g}_{1}\) and the diffusion function \(\mathbf{g}_{2}\) on the right-hand side of Eq.\eqref{ngetg2} using neural networks, such that the model output matches the data-derived quantities.

    The basic structure of neural network (NN) adopted in this project is fully-connected Multi-Layer Perceptrons (MLPs) containing 2 hidden layers and 128 neurons per layer. Two networks are built, one for the drift functions \(\mathbf{g}_1^*(E,U)=(g^*_{11}(E,U),g^*_{12}(E,U))\), and one for random diffusion function \(\mathbf{g}_2^*(E,U)=(g_{21}^*(E,U),\allowbreak g_{22}^*(E,U))\).
    The input dimension is 2 for \((E,U)\).
    The star mark in the output stands for predicted value.
    The network structure is plotted in \figref{NN.fig}. The two networks are connected sequentially, with the output of the drift function network \(\mathbf{g}_1^*(E,U)\) used as a term in the loss function during the training of the diffusion network. This coupling introduces a dependency between the two networks, ensuring their learned dynamics remain consistent.
    
    If we represent \(\mathbf{g}_1\) and \(\mathbf{g}_2\) directly with neural network, the trajectories generated by our model may exhibit non-physical behavior when \(E\) and \(U\) are close to zero, even resulting in negative values of \(E\) and \(U\). For example, in theory, the relationship between \(dE\) and \(E\) is approximately linear, as turbulence represents an instability that grows exponentially with time. Similarly, the relation between \(dU\) and \(U\) is also linear \cite{miki_novel_2011}. Moreover, since the diffusion function governs the stochasticity in the system, which is primarily driven by turbulence, as \(\mathbf{g}_2\) should vanish when \(E\) is small. To preserve these physical properties, we introduce two modifications to incorporate known physics into the neural network architecture.
    First, a element-wise multiplication with the input \((E,U)\) is applied to the output of Drift network to produce \(\mathbf{g_1}\), to inform network that the drift for the energy of turbulence would be zero when there is no turbulence, and the same case for the energy of zonal flow in the absence of zonal flow. Second, output \(\mathbf{g_2^*}\) is obtained by the output of Random diffusion network multiplied by \((E,E)\): randomness comes from turbulence, and no turbulence indicates no randomness. Both modification regulated the behavior for small $E$ and $U$. 
    
    Suppose that after removing buckets with insufficient data, there are $N_b$ valid buckets in total, with the number of data in the $i$-th buckets being $N_i$.
    We also denote the data belonging to the $i$-th box after taking the finite difference in time as $\Delta \mathbf{x}_{i,j}$,
    where $j$ is the index of the data point within the box.
    Following Eq.\eqref{ngetg1} and Eq.\eqref{ngetg2}, the loss for training the drift and diffusion networks take respectively the form of
    \begin{equation}\label{subloss}
        \begin{aligned}
            &L_{D_{\mathbf{g}_1}} = \frac{1}{N_b} \sum_{i=1}^{N_b} 
            \left| \text{UT}\left[\mathbf{g}_1^{*}(\mathbf{x}_i;\theta_1)\right] - \frac{1}{\Delta t}  \overline{\Delta\mathbf{x}_i} \right|^2 \\ 
             &L_{D_{\mathbf{g}_2}} = \frac{1}{N} \sum_{i=1}^{N_b} 
            \Bigg| \text{UT}\left[\mathbf{g}_2^{*2}(\mathbf{x}_i;\theta_2)\right] 
            - \frac{1}{N_i \Delta t} \sum_{j=1}^{N_i} (\Delta \mathbf{x}_{i,j}- \mathbf{g}_{1}(\mathbf{\xi}_i) \Delta t)^2+K_i\Bigg|^2\\
        \end{aligned}
    \end{equation}
    where
    \begin{equation}
        \overline{\Delta\mathbf{x}_i} = \sum_{j=1}^{N_i}\Delta \mathbf{x}_{i,j},
    \end{equation} 
    i.e. the mean value of $\Delta x_{i,j}$ in the $i$-th box, and 
    \begin{equation}
    K_i =
            \begin{cases}
                \text{UT}\left[\mathbf{g}_1^{*2}(\mathbf{x}_i;\theta_1)\right] \Delta t- \mathbf{g}_1^{*2}(\mathbf{\xi}_i;\theta_1)\Delta  t & \text{if } N_i \geq N_t  \\
                0  & \text{if } N_i < N_t
            \end{cases}.
            \label{eq:correction_K}
    \end{equation}
    We note that the approximation $E_\mathbf{x}[\mathbf{g}_1(\mathbf{x})]\approx\mathbf{g}_{1}(\mathbf{\xi}_i)$ is used to derive $L_{D_{\mathbf{g}_2}}$ to simplify the expression.
    The term Eq.\eqref{eq:correction_K} is a correction to take into account the effect of non-zero box width on $\mathbf{g}_1$ as shown in Eq.\eqref{ngetg2}.
    However, when the number of points is not large enough within the bucket, statistic errors will lead to a negative value of $\mathbf{g}_2^2$.
    To avoid this problem, the $K_i$ term is only activated when the number of points within the bucket is above a given threshold $N_t=5000$.

    The training process takes two stages.
    First, only $L_{D_{\mathbf{g}_1}}$ is minimized and the network for $\mathbf{g}_1$ is trained.
    After that, the $\mathbf{g}_1$ network with fixed weights is then used to train the network for $\mathbf{g}_2$.

\subsection{Covariance matrix}

With the learned model, we now obtain \(\mathbf{g}_{1}(\mathbf{x}; \theta_1)\) and \(\mathbf{g}_{2}(\mathbf{x}; \theta_2)\) in equation \eqref{eq10}. Now we need to examine the Brownian motion \(d\mathbf{w}\), specifically the relationship between its two components \(dw_1\) and \(dw_2\).

\begin{figure}[hbt!] 
        \centering
        \includegraphics[width=0.7\textwidth]{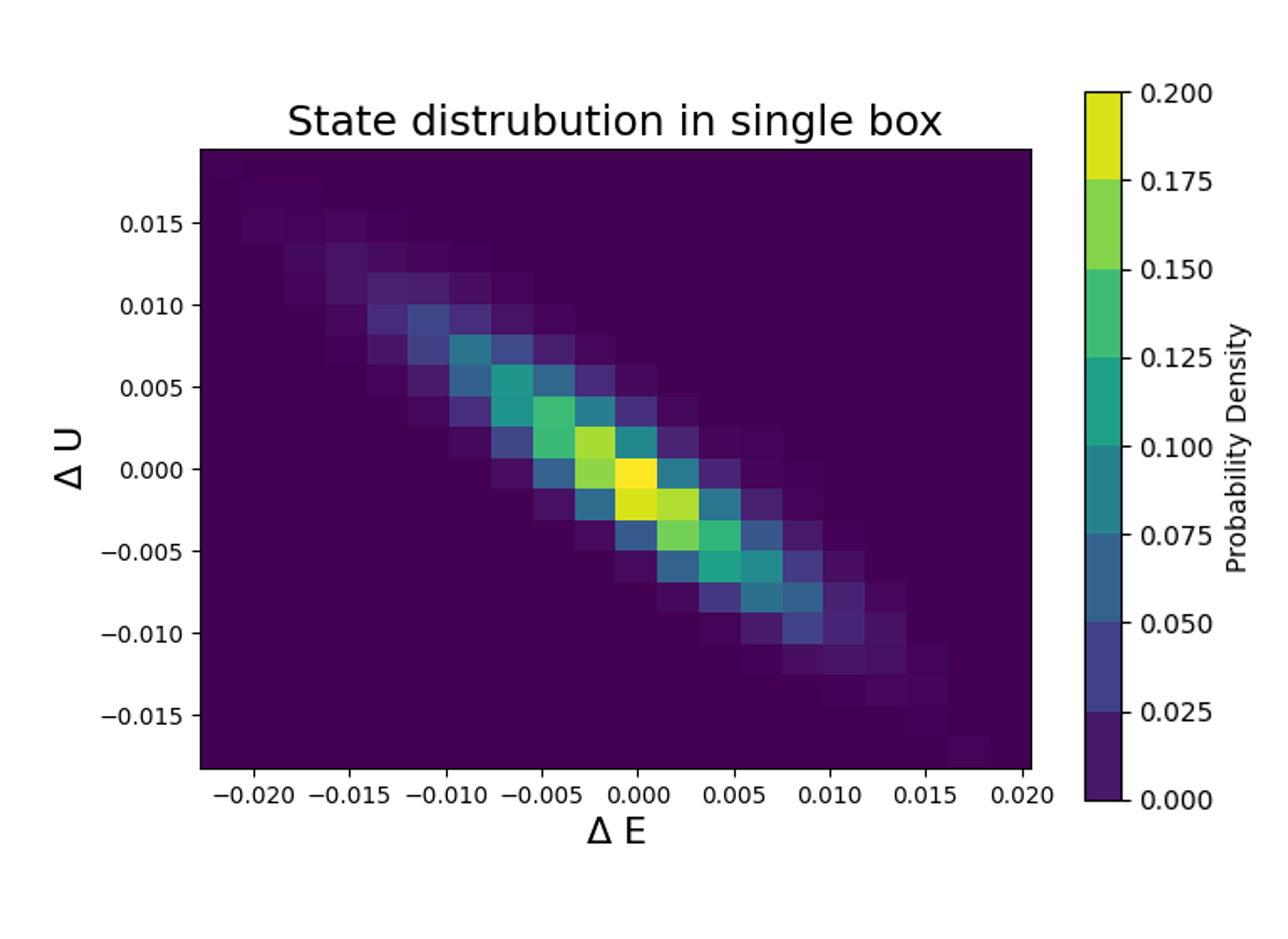} 
        \caption{Distribution of $(\Delta E, \Delta U)$ within a bucket centered at $(E,U)=(0.4,0.4)$ with box width $\rho=0.025$.}
        \label{g2dis}
    \end{figure}

The covariance matrix $\mathbf{\Sigma}$ of \(d\mathbf{w}\) needs to be determined before simulations of the extracted SDEs can be made.
We have analyzed the distribution of $\Delta \mathbf{x} = (\Delta E, \Delta U)$ within each bucket of data.
Within almost all the buckets, there is a strong negative correlation between $\Delta E$ and $\Delta U$ shown in Figure \ref{g2dis}, an example of the distribution in the bucket centered at $(E,U)=(0.4,0.4)$ with box width $\rho=0.025$. The non-diagonal terms in the covariance matrix are found to be around $-0.92$.
This makes physical sense as the total energy of turbulence and zonal flows is approximately conserved: the random fluctuation of one is at the cost of the other.
Therefore, throughout the rest of the paper, we set the covariance matrix to be 
\begin{equation}
    \mathbf{\Sigma} = 
    \left(
    \begin{array}{cc}
       1  & -1 \\
       -1  & 1
    \end{array}
    \right).
\end{equation}
Practically speaking, when we simulate a given set of two SDEs,
at each time step we draw one random number $\nu$ from $\mathcal{N}(0,1)$, 
and apply it to the equation of $\Delta E$ and $\Delta U$ with a positive sign and negative sign, respectively.
    
\subsection{Evaluation of the model}
     Typical machine learning approaches evaluate results based on training and evaluation loss.
     This method is less reliable for SDEs. A low loss value does not necessarily indicate a good capture of the data's features due to stochasticity and limited data availability.
     Instead, we evaluate our model qualitatively and quantitatively in the following two ways.
    
    \subsubsection{Stagnation in predator-prey trajectories}
    
        \indent For a stochastic predator-prey system, each run would produces a difference trajectory even though the initial conditions are the same. However, a common feature is a stagnation~\cite{yoshida_rapid_2003} of the trajectory of turbulence energy and that zonal flow energy, as illustrated in \figref{SDEs1.png}, where two trajectories seem constrained, oscillating within a small range.
        Physically, it represents the predator-prey system has reached an equilibrium point - an attractor. The only way to get out of the attractor is via small fluctuations.
        Whether this common feature appears in predicted trajectories could be a criterion for evaluation.
     
    \subsubsection{Kullback–Leibler (KL) divergence}
    \label{sec:KL}
        The KL divergence measures how a probability distribution \(Q(\mathbf{x})\) diverges from a reference probability distribution \(P(\mathbf{x})\), defined as
        \begin{equation}\label{KLeval}
             D(P|Q) = \int P(\mathbf{x})\log\frac{P(\mathbf{x})}{Q(\mathbf{x})}d{\mathbf{x}}.
        \end{equation}
        A small KL divergence indicates two distributions are similar to each other.
        Here  in our study is the probability distribution for observed states in the mHW simulations, computed from the histogram of $\mathbf{x}=(E,U)$.
        The distribution \(Q(\mathbf{x})\) is similarly obtained by simulating the learned SDEs $50$ times for $10000$ steps, starting from same initial condition as simulations.

\section{Physics-informed extracting process}\label{Result}
\subsection{Detail of training}
    During training, we use the Adam optimizer with a learning rate of 0.001, a decay rate of 0.9, and a decay step of 2000. The model is trained for 40,000 steps with a batch size of 128 on a NVIDIA A100 80GB PCIe GPU, requiring approximately 10 minutes. Additionally, gradient normalization is applied every 1000 steps.
    \subsection{Physically-unconstrained neural network }
\begin{figure}[hbt!] 
            \centering
            \includegraphics[width=1.0\textwidth]{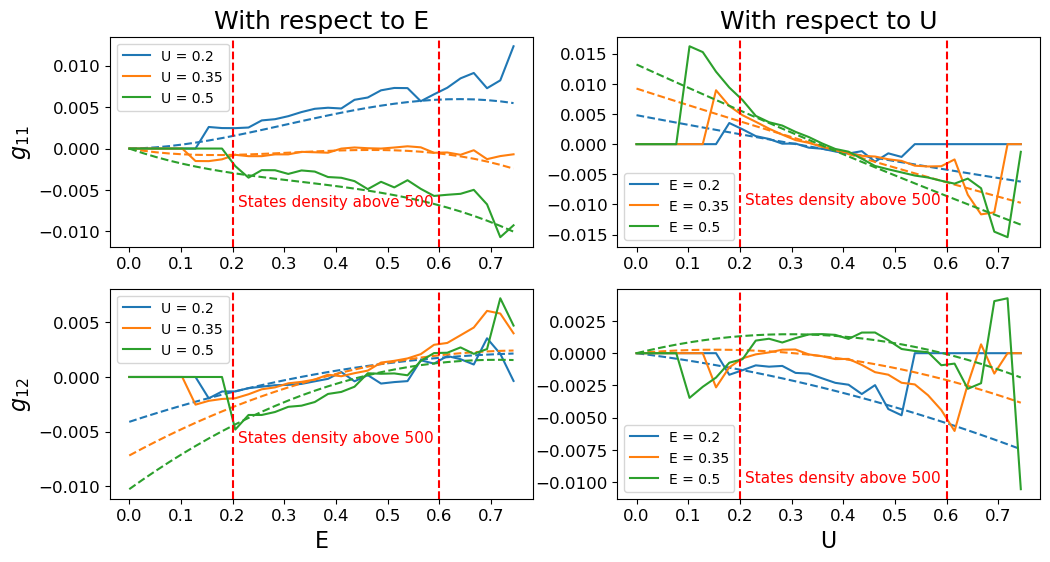} 
            \caption{The value of \(\mathbf{g_1}(E,U)\) elements from the neural network without physics constraints (dashed line), compared to \(\mathbf{g_1}(E,U)\) elements extracted from TOKAM2D simulations (solid line).}
            \label{1Dcomp_old1.fig}
        \end{figure}

\begin{figure}[hbt!] 
            \centering
            \includegraphics[width=1.0\textwidth]{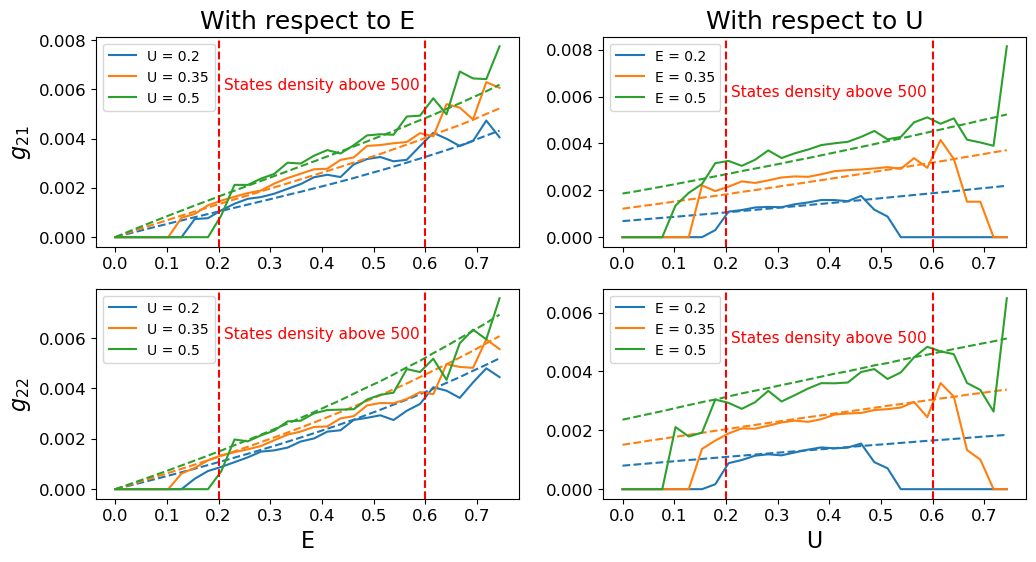} 
            \caption{The value of \(\mathbf{g_2}(E,U)\) elements from the neural network without physics constraints (dashed line), compared to \(\mathbf{g_2}(E,U)\) elements extracted from TOKAM2D simulations (solid line).}
            \label{1Dcomp_old2.fig}
        \end{figure}

\begin{figure}[h] 
            \centering
            \includegraphics[width=0.95\textwidth]{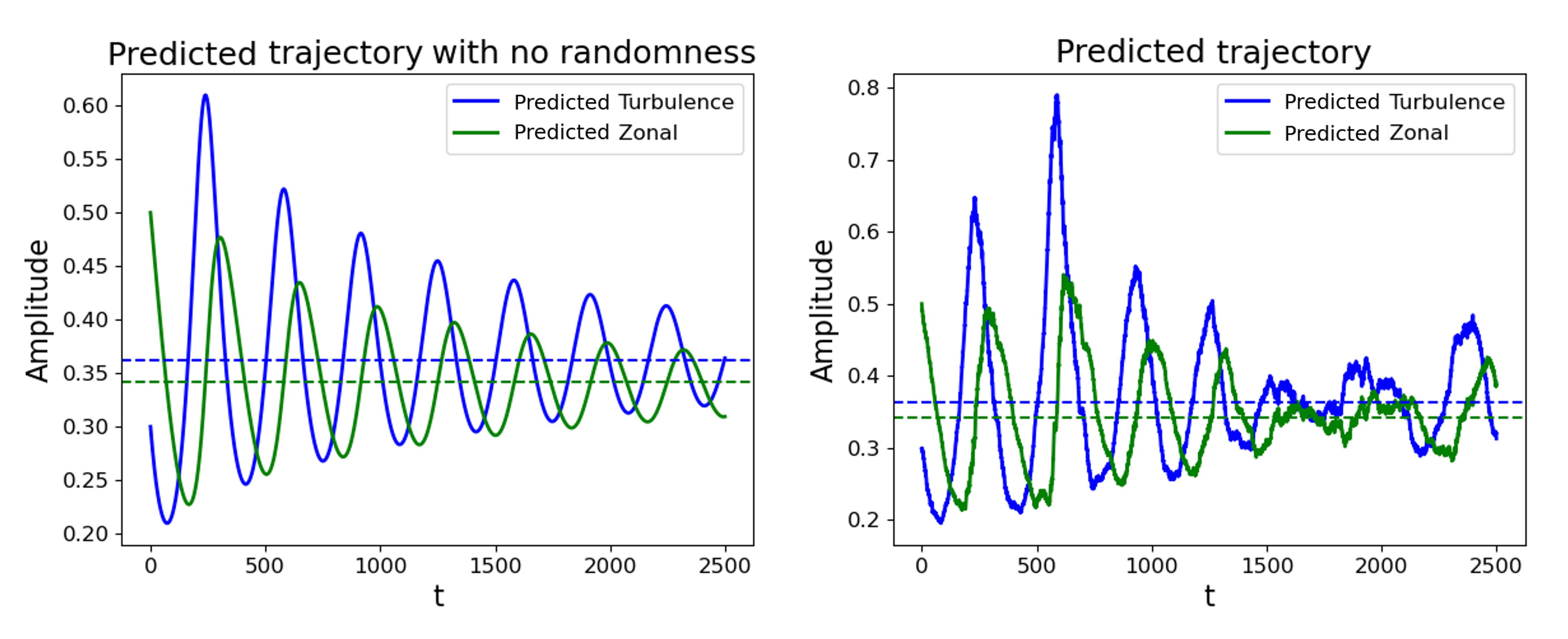} 
            \caption{Time evolution of E and U predicted by the learned neural-network SDE without physics constraints when the stochasticity term is disabled (left) and enabled (right). The dashed lines indicates the averaged level for $E$ and $U$ in each case.}
            \label{trajectory.fig}
        \end{figure}

        We start with the neural network following the architecture in Section \ref{sec:NN}.
        After training, four model functions (\(g_{11}(E,U)\), \(g_{12}(E,U)\), \(g_{21}(E,U)\), \(g_{22}(E,U)\) are obtained.
        A direct comparison can be made with the training data, by taking the mean and variance of $\Delta\mathbf{x}$ described in Eq.\eqref{ngetg1} and Eq.\eqref{ngetg2}, as illustrated in \figref{1Dcomp_old1.fig} and \ref{1Dcomp_old2.fig}.
        Our result shows a fair fit with data in area where the number of data points in a grid box is above the threshold: this area is indicated in figure by the dashed vertical lines. 
        Outside these regions, where the data density falls below the threshold, the data curves exhibit strong fluctuations. In some points, the expectation and variance values are zero due to the absence of data points. It is important to note that the neural network's output in these non-data regions should be regarded as extrapolation.
        Inspection of \figref{1Dcomp_old1.fig} shows that the function \( g_{11} \) exhibits an approximately linear dependence on \( E \) and \( U \), which is characteristic of a predator-prey system.
        Larger values of \( U \) correspond to smaller values of \( g_{11} \) - this is the zonal flow suppression of turbulence as expected physically.
        Similarly, \( g_{12} \) increases with \( E \) as the Reynolds stress increases with a higher turbulence level, though its dependence on \( E \) and \( U \) has more nonlinear characteristics. 
        The imposed boundary conditions are properly satisfied, with \( g_{11}(E=0,U)=0 \), \( g_{12}(E,U=0)=0 \).
        As shown in Figure \ref{1Dcomp_old2.fig},
        the randomness terms \( g_{21} \) and \( g_{22} \) both increases with a larger $E$ and $U$, though the dependency on $E$ is stronger than on $U$.
        The two terms show a strong correlation, displaying similar trends and magnitudes.
        The boundary conditions are also satisfied:
        both \( g_{21} \) and \( g_{22} \) to be zero when \( E=0 \).
        Finally, the magnitude of $\mathbf{g}_1$ and $\mathbf{g}_2$ is similar.
        It reveals the intrinsic difficulty of learning the turbulence-zonal flow dynamics directly from data: the randomness is as important as the drift in determining the trajectories, or in other words, the noise is multiplicative and its ratio is nearly or even higher than 100\%\cite{Robin_2025}.
        With learned drift functions and diffusion functions, we can study the predator-prey trajectories without randomness.
        The predator-prey trajectories of $(E,U)$ without randomness are obtained by only enabling the time drift terms \(\mathbf{g}_1\), as shown in \figref{trajectory.fig}.
        There is an obvious damping effect as the range of the predator-prey oscillation is shrinking with time and their value converging into a pair of constants (an attractor), represented by the parallel dash lines in \figref{trajectory.fig} and they are obtained by taking average of states over the trajectory.
        While the $\mathbf{g}_2$ term is enabled, even though the amplitude of the predator-prey oscillation still damps initially, it rises again after spending some time around the attractor.
        It suggests that the randomness acts as a drive in our extracted the SDEs predator-prey system of turbulence and zonal flow, since the amplitude of the predator-prey oscillation from TOKAM2D also suggests no sign for damping (see, for example, \figref{SDEs1.png}).
        This effect explains the stagnation observed in TOKAM2D data and SDE prediction, because in our model, the state would be stuck in the attractor without randomness, but random fluctuations play a role in pushing the state out of the attractor. And it takes some time for randomness to push the state away from the attractor, because random movement needs to accumulate to counter the gravity of the attractor.
        
\begin{figure}[hbt!] 
            \centering
            \includegraphics[width=1.0\textwidth]{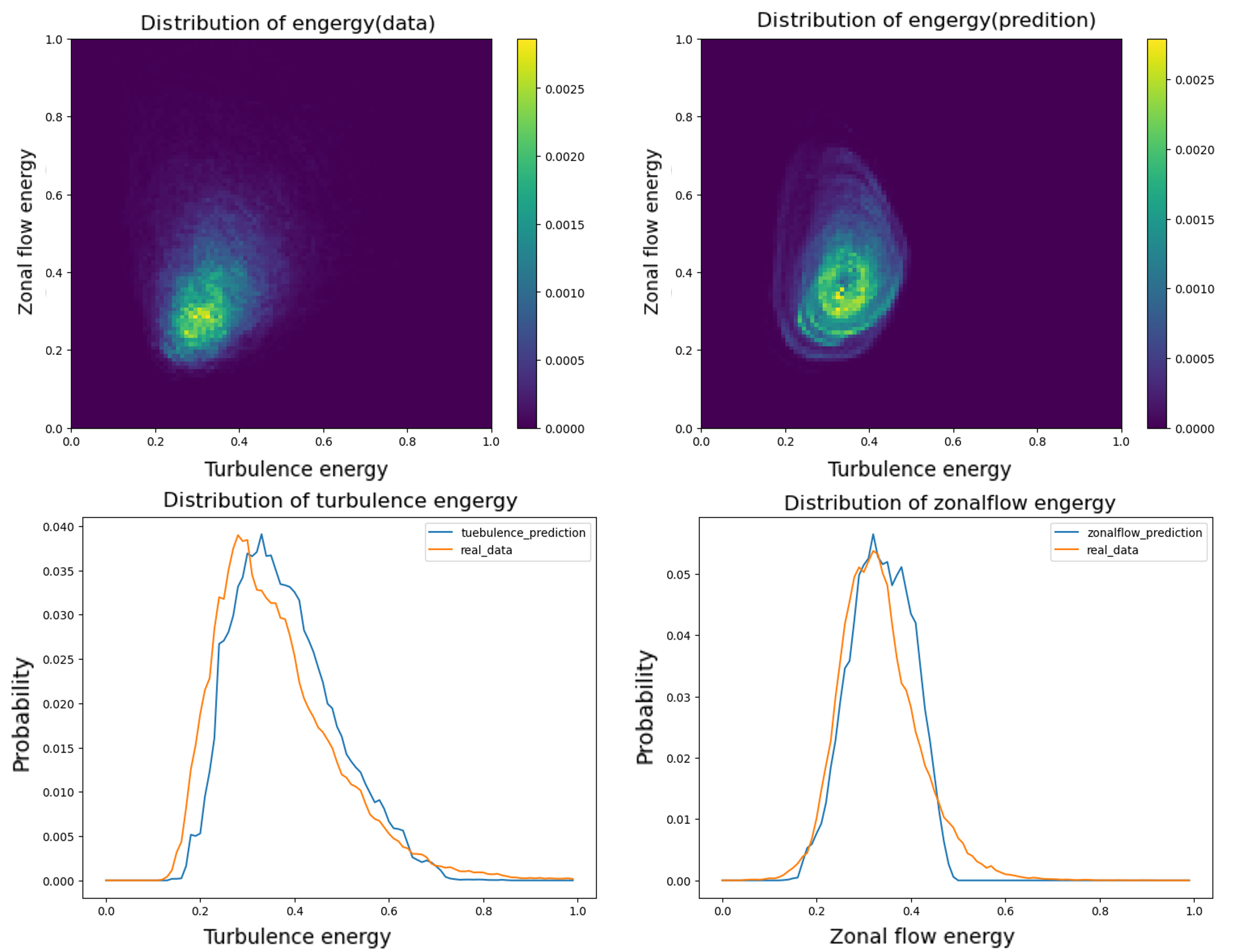} 
            \caption{Predicted states distribution from SDEs model without physics constraints, compared to the distribution from TOKAM2D data.}
            \label{density.fig}
        \end{figure}

        To compare the distribution of $(E,U)$ and compute the KL divergence, we generated 50 SDE simulations of 10000 time steps as described in Section \ref{sec:KL}, with the results shown in \figref{density.fig}.
        The two-dimensional distributions from the TOKAM2D data and the learned SDE are very similar.
        They both have a finger-print shape centered around $(E,U)=(0.35,0.35)$, the location of the equilibrium point, except the presence of spurious rings in the peripheral region for the SDE results.
        Examining the one-dimensional distributions, the model generally shows fair agreement with the data. However, closer inspection reveals that the match deteriorates near the boundaries. This discrepancy is expected: neural networks perform best when trained on abundant data, whereas peripheral regions often contain few or no data points. These peripheral inaccuracies can cause the simulated trajectories to behave strangely near the edges of the state space, leading to mismatches in those regions.
        To compare the two distributions, we also evaluate their KL divergence between them. The KL divergence is computed using equation \eqref{KLeval}, based on the distribution functions of \((E, U)\) obtained from the TOKAM2D data and the simulations generated by the learned SDE model.
        Since each group of SDE simulations generates a slightly different distribution due to randomness, we repeat the process 10 times to compute the mean and standard deviation. 
        The value of KL divergence is $0.48\pm0.02$: an acceptable value, though some improvements can be made.
        The main contribution to the KL divergence comes from the peripheral regions where the ratio of $P(\mathbf{x})/Q(\mathbf{x})$ can be very large for $Q(\mathbf{x})\rightarrow0$.
        it is therefore possible to obtain a better KL divergence by improving the accuracy of the extrapolation to the region with insufficient data.


        One way to improve the accuracy of extrapolation is to construct an analytical formula from the learned networks $\mathbf{g}^*_1$ and $\mathbf{g}^*_2$ within the region with good statistics, then use it for extrapolation.
        We fit the analytical formula to the output of the neural network, with details given in \ref{app:maths_form}.

        By extracting an analytical formula from the model, we eliminate the presence of spurious rings in the peripheral regions of the two-dimensional distribution, resulting in a shape that more closely matches the data. However, the extracted formula introduces a shift in the distribution's center, which leads to a suboptimal KL divergence value despite the improved overall agreement in shape. 
        The shift in the center of the 2D is caused by errors introduced during the extraction of the analytical formula. In fact, the distribution of states is highly sensitive to the detailed shape of the extracted functions. Even a 2\% difference in the coefficients can lead to significant changes in the resulting distribution. As a result, even small inaccuracies in the analytical formula can cause a noticeable shift in the distribution center.

        Finally, it is noteworthy that using the standard linear regression method or SINDY\cite{champion_data2019} directly on data would get a messy mathematical form with a high error for \(\mathbf{g_1}\), primarily due to the strong stochasticity inherent in the our model. These directly extracted mathematical forms can lead to unphysical behavior in simulations. For example, they may cause explosion in the simulated SDE trajectories, where \(E\) and \(U\) exceed 1 or fall below 0. Furthermore, the predicted deterministic trajectories (i.e., with no randomness) show suppressed predator–prey dynamics, where \(E\) and \(U\) fail to exhibit the expected oscillations.

    \subsection{Adding more physical constraints}\label{subr}

        The result presented in the last section reveals a problem.
        On one hand, the distribution of states is very sensitive to the detailed shape of $\mathbf{g}_1$ and $\mathbf{g}_2$.
        On the other hand, due to the high randomness in the system, the fluctuations in the data and the variance in the fitting are large, i.e. there are a lot of different ways to fit the same data with same fitting error and it is difficult to distinguish which one is better.
        The model is therefore prone to overfitting without a strong regularization.
        So far, we are relying on the bucket-averaging method and the regularization within the neural network itself to reduce variance.
        We can, instead, use more physics-based knowledge to regularize our fitting, therefore reduce the arbitrariness of the learned neural network.
        
\begin{figure}[hbt!] 
            \raggedright
            \includegraphics[width=1.\textwidth]{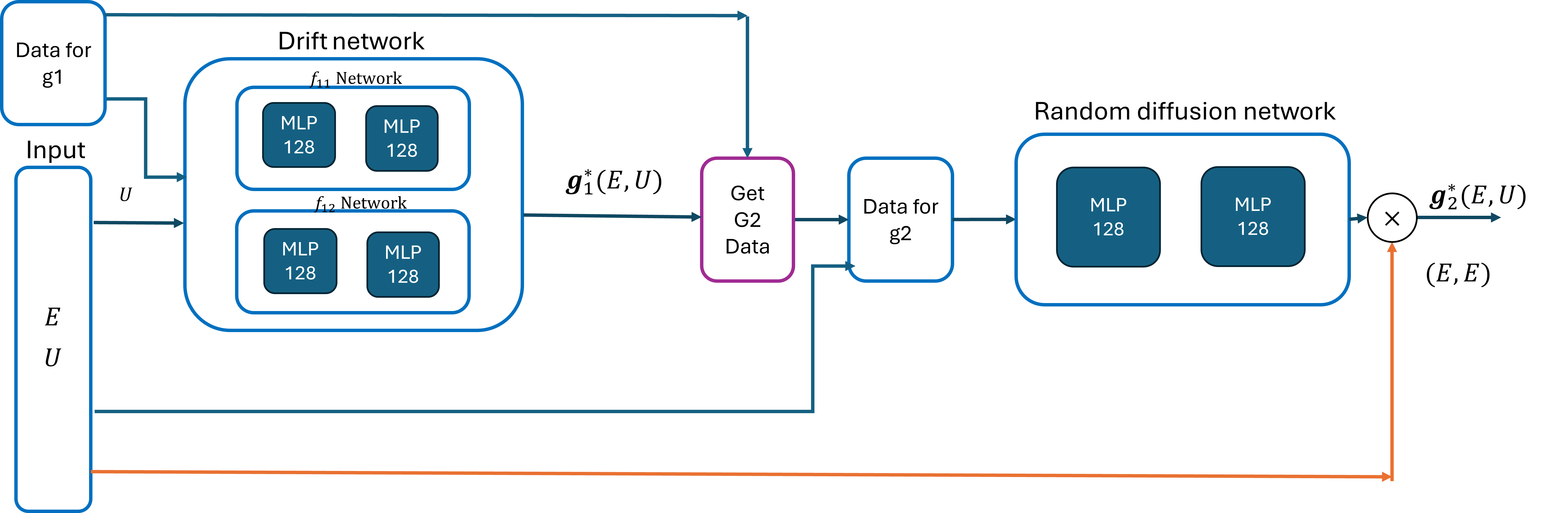} 
            \caption{New NN structure with more physics constraints, where \(E\) and \(U\) represent the energy of turbulence and the energy of zonal flow respectively. }
            \label{NN1.fig}
        \end{figure}

\begin{figure}[hbt!] 
            \centering
            \includegraphics[width=1.0\textwidth]{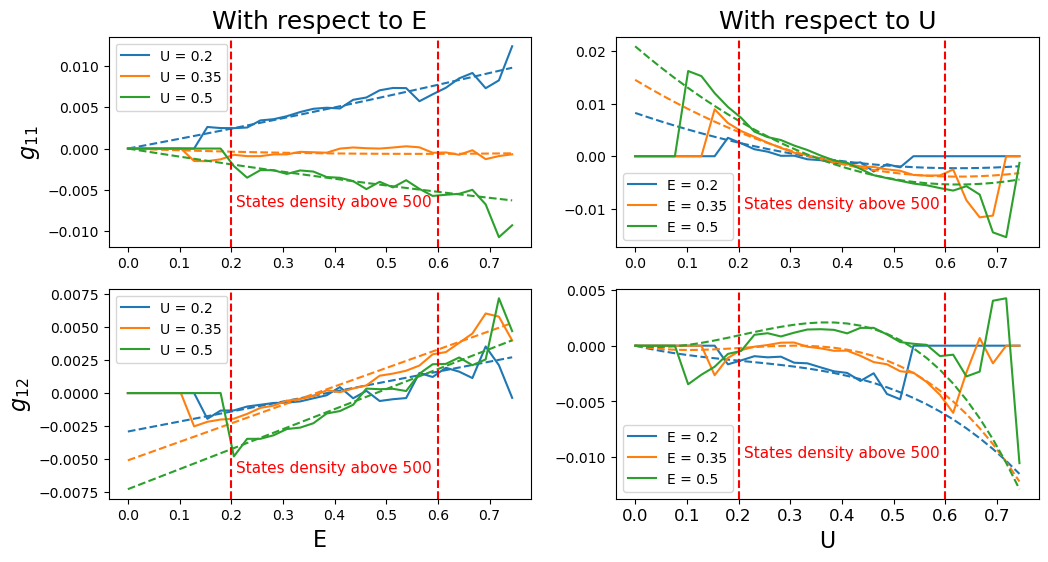} 
            \caption{The value of \(\mathbf{g_1}(E,U)\) from the neural network with physics constraints (dashed line), compared to \(\mathbf{g_1}(E,U)\) extracted from TOKAM2D simulations (solid line).}
            \label{1Dcomp1.fig}
        \end{figure}

\begin{figure}[hbt!] 
            \centering
            \includegraphics[width=1.0\textwidth]{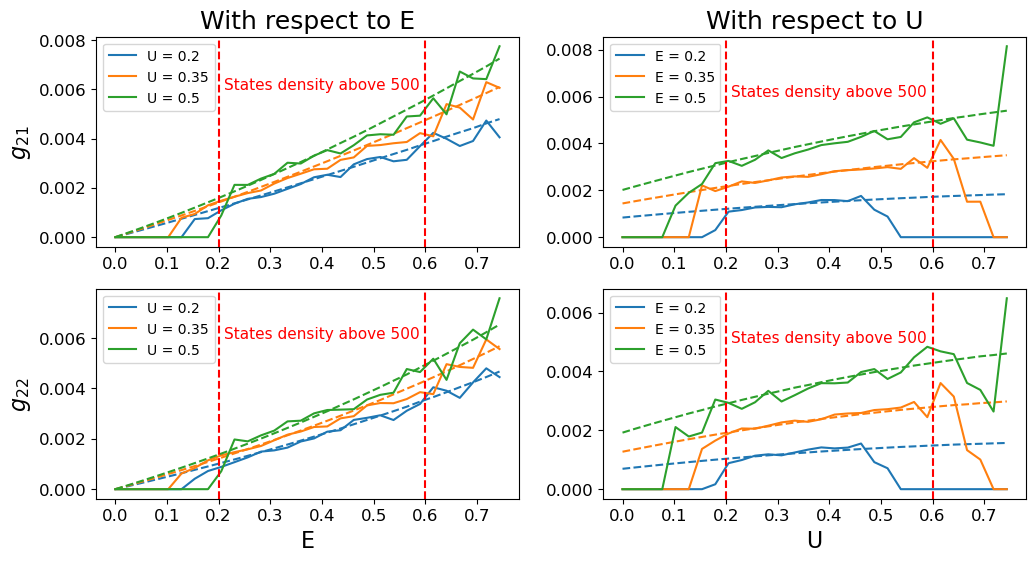} 
            \caption{The value of \(\mathbf{g_2}(E,U)\) from the neural network with physics constraints (dashed line), compared to \(\mathbf{g_2}(E,U)\) extracted from TOKAM2D simulations (solid line).}
            \label{1Dcomp2.fig}
        \end{figure}

\begin{figure}[hbt!] 
            \centering
            \includegraphics[width=1.0\textwidth]{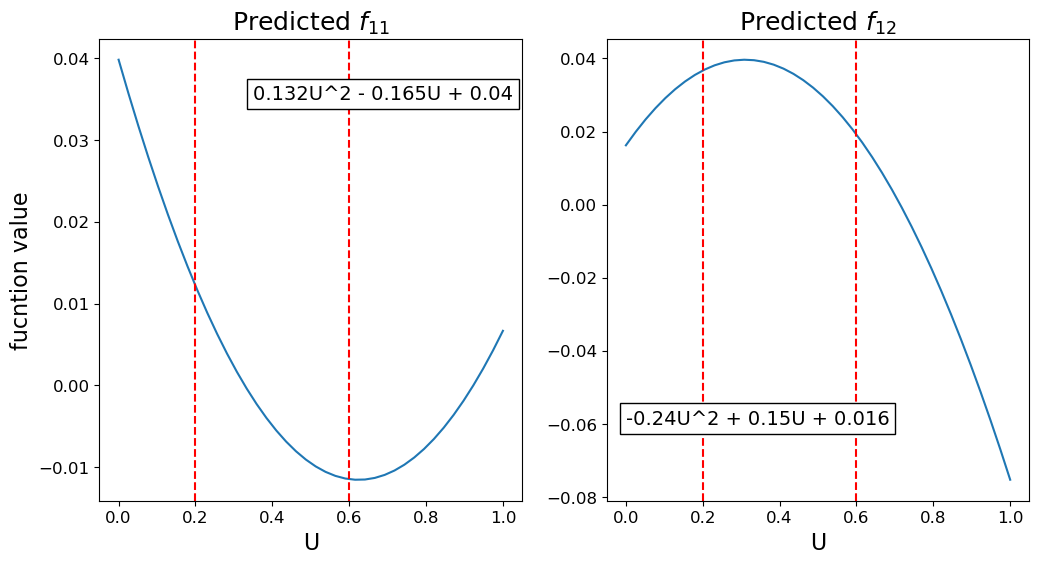} 
            \caption{Learned \(f_{11}\) and \(f_{12}\) as function of $U$. Two red vertical lines enclose an area with sufficient data.}
            \label{f1.fig}
        \end{figure}

        \indent Instead of letting $\mathbf{g}_1$ to be an arbitrary function of $(E,U)$, we constrain its form to follow
        \begin{equation}\label{HWc}
            \begin{aligned}
                g_{11}(E,U,\theta_1)&= Ef_{11}(U,\theta_1)+AE^2, \\
                g_{12}(E,U,\theta_2)&= - BU + E U f_{12}(U,\theta_2),
            \end{aligned}
        \end{equation}
        where \(f_{11}(U,\theta_1)\) and \(f_{12}(U,\theta_2)\) are two neural networks taking only the zonal flow energy $U$ as input with their parameters \(\theta_1\) and \(\theta_2\), respectively. The constants \(A\) and \(B\) are learnable parameters, where the coefficient $A$ represents the nonlinear damping of turbulence, while $B$ corresponds to the linear damping rate of the zonal flow.
        We assume that the turbulence self-interacts nonlinearly only through a quadratic term $A E^2$, which does not depend on zonal flow shear.
        The linear growth of $E$ is contained in $f_{11}$.
        We also assume that the zonal flow shear linearly damps through the $-BU$ term, and that the interaction with turbulence is proportion to $E$.
        The additional $U$ in front of $f_{12}$ is to ensure $g_{12}\rightarrow0$ when $U\rightarrow 0$.
        The structure of the new NN for $f_{11}$ and $f_{12}$ is illustrated in \figref{NN1.fig}. The diffusion terms $\mathbf{g}_2$ and the training process are unchanged. \\
        \indent The trend of learned time drift function \(g_{1}(E,U)\) and random diffusion \(g_{2}(E,U)\) function for this new network are depicted in \figref{1Dcomp1.fig} and \ref{1Dcomp2.fig}.
        Again, the match between the neural network and TOKAM2D data is fair within the region with sufficient data - showing that this physically informed step improve the previous results shown in \figref{1Dcomp_old1.fig} and \ref{1Dcomp_old2.fig}. NN captures details that previously cannot be captured, especially for \(g_{11}(E,U)\) and \(g_{12}(E,U)\) where both trends with respect to \(U\) show a slight bending compared with the result before. Those curvatures are captured by \(f_{11}\) and \(f_{12}\), as depicted in \figref{f1.fig}, showing their $U$ dependence. 
        The value of constants \(A\) and \(B\) are found to be $0.002$ and $0.014$, respectively. 
        We will defer the physical interpretation of the results to Section \ref{sec:scan}.
        
        
        \indent  
        \indent \\
        \indent 

\begin{figure}[hbt!] 
            \centering
            \includegraphics[width=1.0\textwidth]{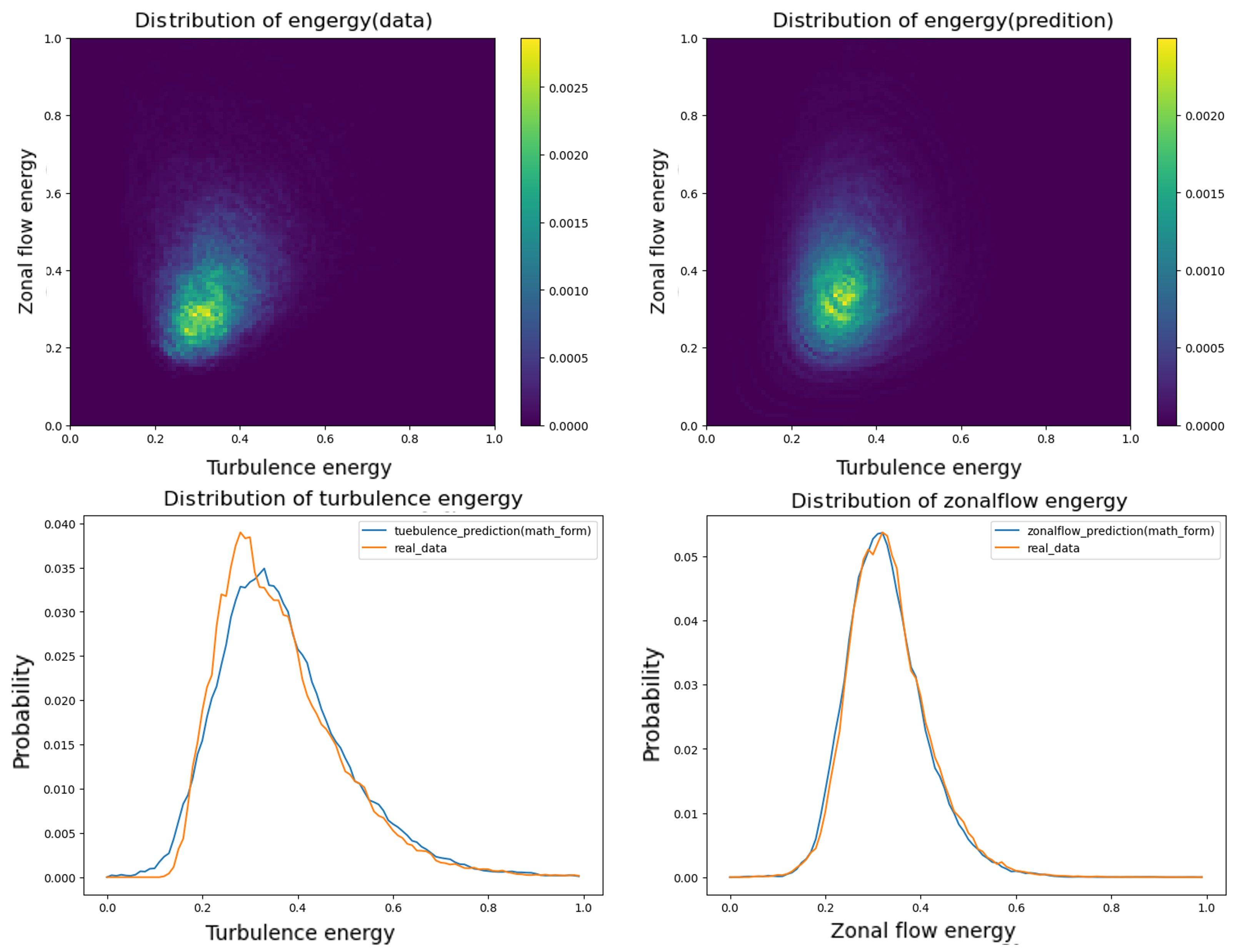} 
            \caption{Predicted states distribution from the learned NN model with physics constraints, compared to the distribution from TOKAM2D data.}
            \label{madis1.fig}
        \end{figure}

\begin{figure}[hbt!] 
            \centering
            \includegraphics[width=0.9\textwidth]{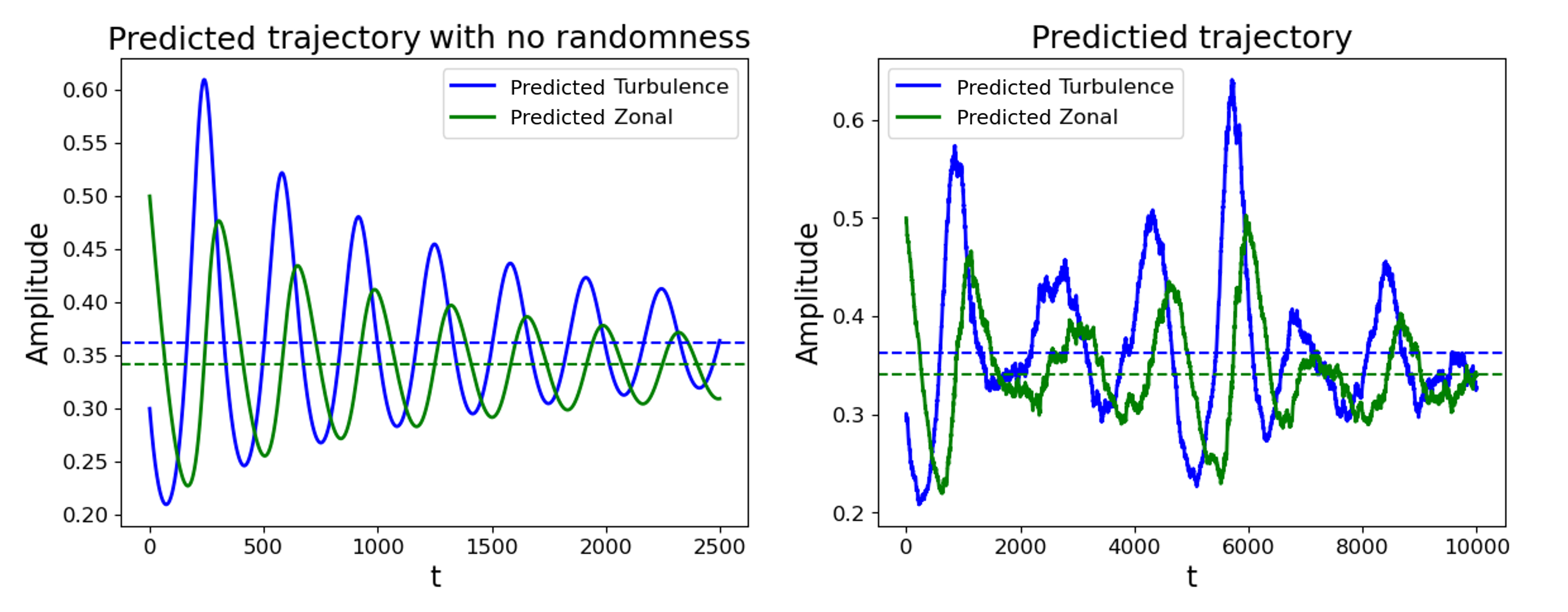} 
            \caption{Simulations by integrating the learned SDE with physics constraints when the stochasticity term is disabled (left) and enabled (right). The dashed lines indicates the averaged level for $E$ and $U$ in each simulation.}
            \label{trajectory_math.fig}
        \end{figure}

        We now fit the output of $f_{11}$ and $f_{12}$ from neural networks into quadratic functions for a better extrapolation into the region with insufficient data.
        The mathematical form of the fitting is given in \figref{f1.fig}.
        Then we generate 50 simulation of 10000 time steps based on the fitted quadratic functions. The state distribution shows a better match with the data as shown in \figref{madis1.fig}. The mean KL divergence between the predicted state density distribution and the data-derived state density reduces to $0.15\pm0.013$, indicating a significant improvement in accuracy. 
        In addition, by observing a single trajectory as illustrated in \figref{trajectory_math.fig},
        the predator-prey oscillation damps in the absence of randomness,
        and the stagnation phenomenon is clearly visible when randomness is introduced. 
        

        In conclusion, embedding physical constraints to the neural network helped to reduce the variances in fitting and lead to better accuracy.
        
\section{Physical interpretation and parameter scan}\label{sec:scan}
\begin{figure}[hbt!] 
            \centering
            \includegraphics[width=0.7\textwidth]{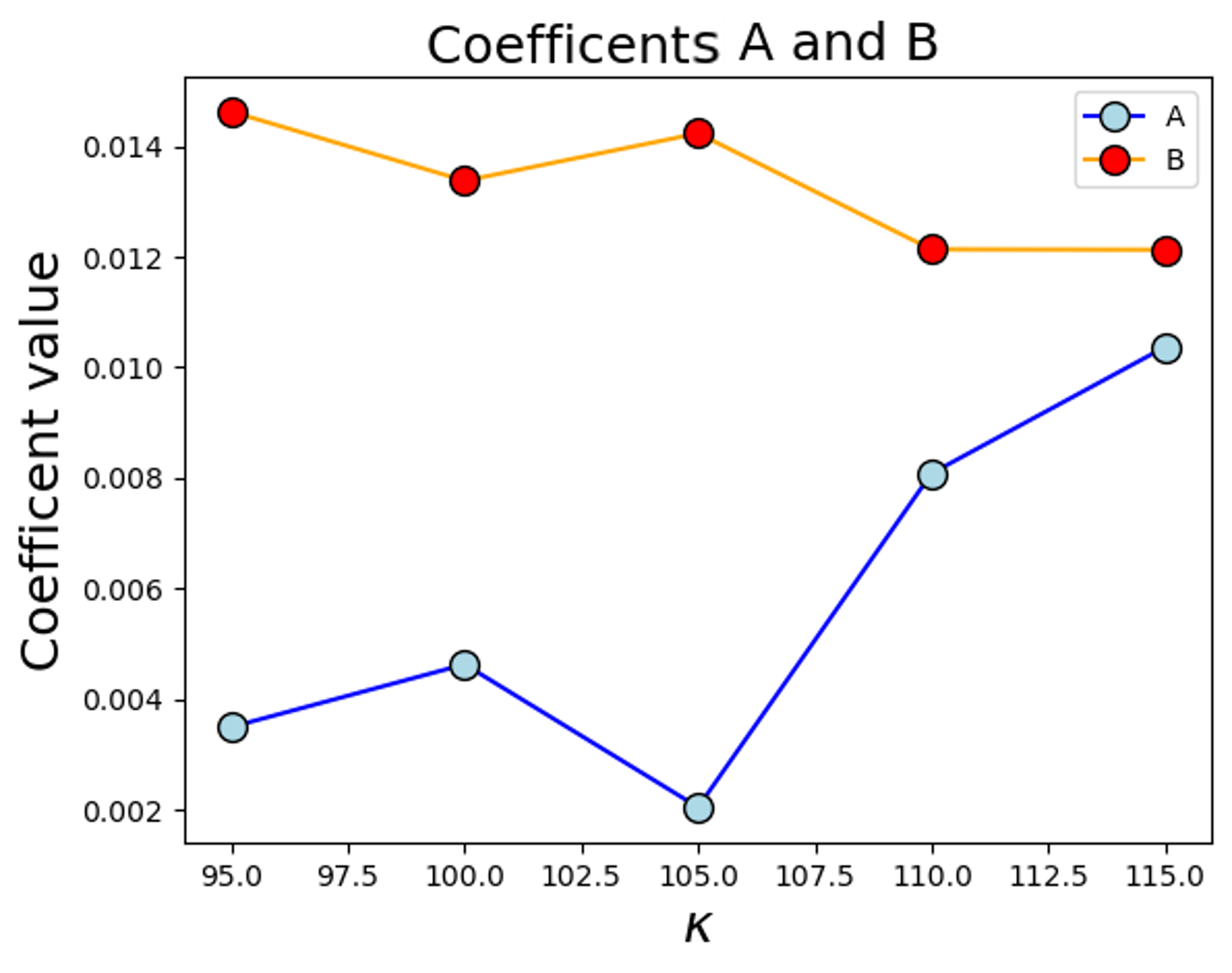} 
            \caption{Coefficients A and B with respect to $K$($K=100\kappa$)}
            \label{coef.fig}
        \end{figure}

        Taking the same process, we performed a parameter scan over the density gradient coefficient $\kappa$ from $1$ to $1.15$ with a step size of $0.05$. This range is chosen because, without increasing the amount of data, a large $\kappa$ leads to a looser state density distribution, which in turn reduce the size of confidence area. For each $\kappa$, we run 50 TOKAM2D simulations with different random seeds. We have extracted the parameters $A$ and $B$, as well as $f_{11}$, $f_{12}$, and $\mathbf{g}_{2}$, using the method in Section \ref{subr}. 
        The exact values and mathematical forms are given in \ref{app:scan_results}.
        For each extracted model, we evaluate the Kullback–Leibler (KL) divergence and examine the presence of stagnation features, as summarized in Table~\ref{tab:KL}. \\
        \begin{table}[h]
          \centering
            \begin{tabular}{|l|c|c|}
            \hline
            $\kappa$ & KL divergency & Stagnation \\
            \hline
            0.95 & $0.12\pm0.01$ & True\\
            \hline
            1.00 & $0.13\pm0.01$ & True\\
            \hline
            1.05 & $0.15\pm0.013$ & True\\
            \hline
            1.10 & $0.15\pm0.01$ & True\\
            \hline
            1.15 & $0.17\pm0.01$ & True\\
            \hline
            \end{tabular}
          \caption{KL divergency and Stagnation for each $\kappa$}
          \label{tab:KL}
        \end{table}
        \indent The coefficients $A$, $B$ with respect to the gradient parameter $\kappa$ is shown in Fig.~\ref{coef.fig}. Our model yields a positive value for $A$, whereas theoretical expectations suggest that this term should be negative\cite{diamond_zonal_2005}. One possible explanation lies in a modeling simplification: the stochastic term $dw$ is treated as a normally distributed random variable at each time step. This assumption neglects the temporal continuity of turbulence, as in reality turbulence does not suddenly appear or disappear. As a result, the model underestimates the cumulative effect of stochasticity over time. To compensate for the missing time-accumulated randomness, the model effectively adjusts by assigning a positive value to $A$.
        As for coefficient $B$, it is found to be relatively insensitive to changes in $\kappa$ in this range, which is expected since it primarily originates from the prescribed damping coefficient $\nu$.
        Interestingly, the value of $B$ is less than the theoretical value from linear theory given by $B=2 \nu = 0.02$.
        \\

        The function \(f_{11}\) is the effective growth rate of turbulence considering the presence of zonal flow.
        Turbulence is effectively suppressed by zonal flow since \(f_{11}\) decreases as $U$ increases, until the growth rate becomes negative for $U>0.35$.
        We find that for large $U$, the value of $f_{11}$ is saturated around $-0.01$, indicating limit in efficiency to suppress turbulence.
        Furthermore, while the shape of $f_{11}$ remains largely unchanged across different values of $\kappa$ in the range $[0.95,1.15]$, its overall value shifts up according to the increased linear growth rate. To some extent, it is consistent with a physical facts that the strength of the turbulence–zonal flow interaction depends on the intrinsic properties of the turbulence and zonal flow rather than on the linear growth rate\cite{diamond_self-regulating_1994}.
        This presents the first quantitative confirmation of the theoretical prediction that the efficiency of zonal flow shearing effects on turbulence decreases as the zonal flow energy increases.
    
    
        The physical meaning of $Uf_{12}$ is the efficiency of turbulence in producing Reynold stress, which then drive the zonal flow.
        Similarly, the efficiency decreases with stronger zonal flow.
        It is also observed that $f_{11}$ and $U f_{12}$ exhibit a near mirror symmetry, flipping up side down, which is consistent with energy conservation principles, as energy is exchanged between turbulence and zonal flow through these terms. However, the amplitude of $U f_{12}$ is noticeably smaller than that of $f_{11}$.This discrepancy is possibly due to the assumption that the stochastic components $dw$ of turbulence and zonal flow are negatively correlated.

        \begin{figure}[hbt!] 
            \centering
            \includegraphics[width=0.9\textwidth]{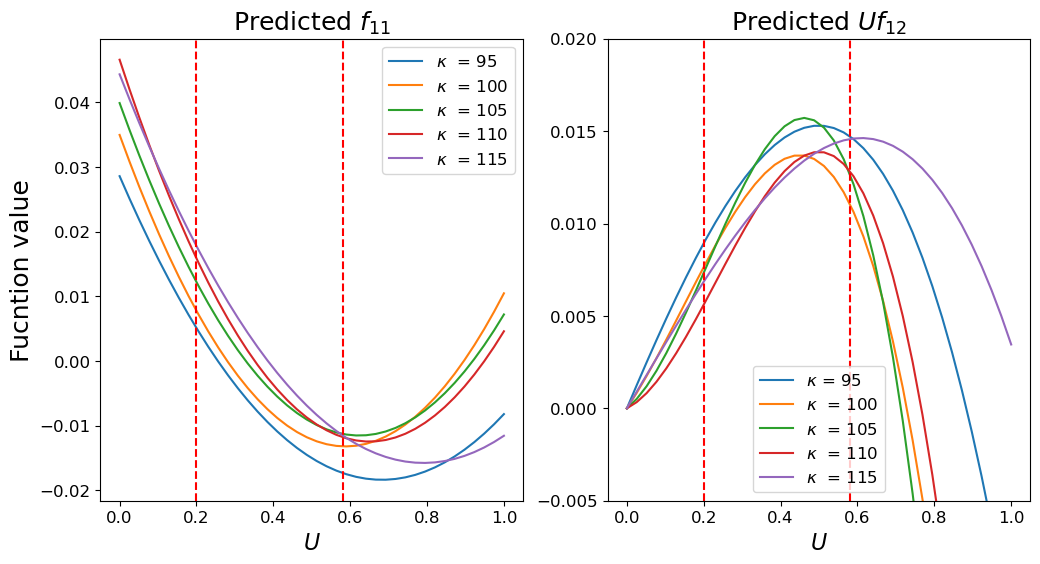} 
            \caption{The learned function $f_{11}$ and $f_{12}$ with respect to $\kappa$, where vertical red dot line enclose the region with sufficient data($K=100\kappa$)}
            \label{f1_math.fig}
        \end{figure}

\section{Conclusion}
    
    We have developed a method to extract a model that captures the predator-prey behavior between turbulence and zonal flow. It is based on neural networks with known physical principles. This approach effectively identifies the key features of the underlying dynamics using a limited dataset, as the SDEs model require a large number of statistical quantities such as expectation and variance. We recover the
    well-known result that the efficiency of zonal flow
    shearing effects on turbulence decreases as the zonal
    flow energy increases. Additionally, we have found that in
    the absence of randomness, predator-prey dynamics between
    turbulence and zonal flows would damp. We have also
    performed a parameter scan over the density gradient
    length, showing that the turbulence-zonal flow interaction is insensitive to the linear growth rate in our scanning range $[0.95,1.15]$.
    
    The resulting model remains relatively simple, though it does not account for the temporal coherence of turbulence, as in reality turbulence does not suddenly appear or disappear. 
    Future work could address this limitation by using stochastic diffusion terms that accumulate in time. In this study, we have also assumed a fixed covariance of $-1$ between the stochastic components of turbulence and zonal flow. A more detailed investigation is required to accurately characterize the correlation between these random fluctuations.
    We have also neglected the zonal density, which is corrugated radially, forming the $E\times B$ staircases.
    Finally, the KL divergence can be targeted as part of the loss function, as mentioned in a recent variant of the SINDY algorithm\cite{jacobs_hypersindy_2023,lenfesty_uncovering_2025}.

\section*{Acknowledgments}
This work is partly funded by Ministry of Education (MOE) AcRF Tier 1 grants RS02/23 and RG156/23, and National Research Foundation Singapore (NRF) core funding ``Fusion Science for Clean Energy''.
The computational work for this article was partially performed on resources of the National Supercomputing Centre (NSCC), Singapore.
    
\section*{References}
\bibliography{references}

\appendix
\section{Unscented transform(UT)}
\label{app:UT}
    The Unscented Transform (UT) is a technique used to estimate the statistics of a nonlinear transformation of a random variable, particularly when the distribution of the variable is approximated by a Gaussian distribution. \\
    \indent Suppose that one wants to measure the moment of a random system \(\mathbf{y} = F(\mathbf{z})\), where \(\mathbf{z}\) denotes state. The mean and covariance matrix of states for such system are $m$ and $P$. Suppose \(\mathbf{z}\) is an n-dimension vector, Unscented transform is enforced as follows:\\
    1. Form \(2n+1\) sigma points by combining knowledge of the mean and covariance matrix:

    \begin{equation}
        \begin{aligned}
            \mathbf{z}^{(0)} &= \mathbf{m}, \\
            \mathbf{z}^{(i)} &= \mathbf{m} + \left[\sqrt{(n + \lambda) \mathbf{P}}\right]_i, \quad i = 1, 2, \ldots, n ,\\
            \mathbf{z}^{(j)} &= \mathbf{m} - \left[\sqrt{(n + \lambda) \mathbf{P}}\right]_j, \quad j = n+1, n+2, \ldots, 2n,
        \end{aligned}
    \end{equation}
    ,where \(\left[\sqrt{(n + \lambda) \mathbf{P}}\right]_i\) denotes the \(i\)-th column of the Cholesky decomposition of \((n + \lambda) \mathbf{P}\). Here, \(\lambda = \alpha^2 (n + \kappa) - n\), and \(\alpha\in(0,1)\) and \(\kappa\) are constants.\\
    2. Form \(2n+1\) weights for these sigma points:
    \begin{equation}
        \begin{aligned}
            W_0^{(c)} &= \frac{\lambda}{n + \lambda} - (1 - \alpha^2) ,\\
            W_0^{(m)} &= \frac{\lambda}{n + \lambda} ,\\
            W_i^{(c)} &= \frac{1}{2(n + \lambda)}, \quad i = 1, 2, \ldots, n ,\\
            W_j^{(m)} &= \frac{1}{2(n + \lambda)}, \quad j = n+1, n+2, \ldots, 2n,
        \end{aligned}
    \end{equation} here one could modify weight by tuning \(\lambda\) and \(\alpha\), for example the larger \(\lambda\) means the heavier weight on mean \(\mathbf{m}\),  larger \(\alpha\) means small weight on mean \(\mathbf{m}\).  \\
    3. Apply function \( F \) to these sigma points:
    \begin{equation}
        y^{(i)} = F(\mathbf{z}^{(i)}), \quad i = 0, 1, \ldots, 2n
    \end{equation}
    4. Get the mean and covariance of \( y \):
    \begin{equation}
        \hat{\mu}_y = \sum_{i=0}^{2n} y^{(i)} W_i^{(m)}
    \end{equation}
    \begin{equation}
        \hat{\Sigma}_y = \sum_{i=0}^{2n} (y^{(i)} - \hat{\mu}_y)(y^{(i)} - \hat{\mu}_y)^T W_i^{(c)}
    \end{equation}

\section{Maths form extracted from neural network}
\label{app:maths_form}
        The simplest non-trival form motivated by \figref{1Dcomp_old1.fig} and \figref{1Dcomp_old2.fig} is given by
        \begin{equation}\label{mathform}
            \begin{aligned}
                &dE = (a_1 E +b_1 EU + c_1 E^2)dt+(d_1 E+e_1 EU)dw,\\
                &dU = (a_2 U +b_2 EU + c_2 U^2)dt-(d_1 E+e_1 EU)dw,
            \end{aligned}
        \end{equation} 
        where \(a_{1,2}\), \(b_{1,2}\), \(c_{1,2}\), \(d_1\) and \(e_1\) are constant coefficients.

        To get math expression, we fit predicted function in area where data density above 500, and find functions take forms as \(dE = (0.009E + 0.004E^2 - 0.038EU)dt+(0.008E + 0.008 EU )dw\),  \((-0.005U - 0.005U^2 + 0.026EU)dt - (0.008E + 0.008 EU )dw\) with a linear regression error \(0.044\). One thing need to note is that terms $0.004E^2$ and $0.005U^2$ in these function play a significant role, even their magnitudes are small(take about $2\%$ less of total function value in confident area): For our data, if we remove these terms from function \eqref{mathform} to get a new function, however, the latter form results in a significantly larger KL divergence and more likely to cause explosion in simulation of trajectory. This indicates that the stochastic system is highly sensitive, even to small changes in its governing dynamic functions.
        \begin{figure}[hbt!] 
            \centering
            \includegraphics[width=1.0\textwidth]{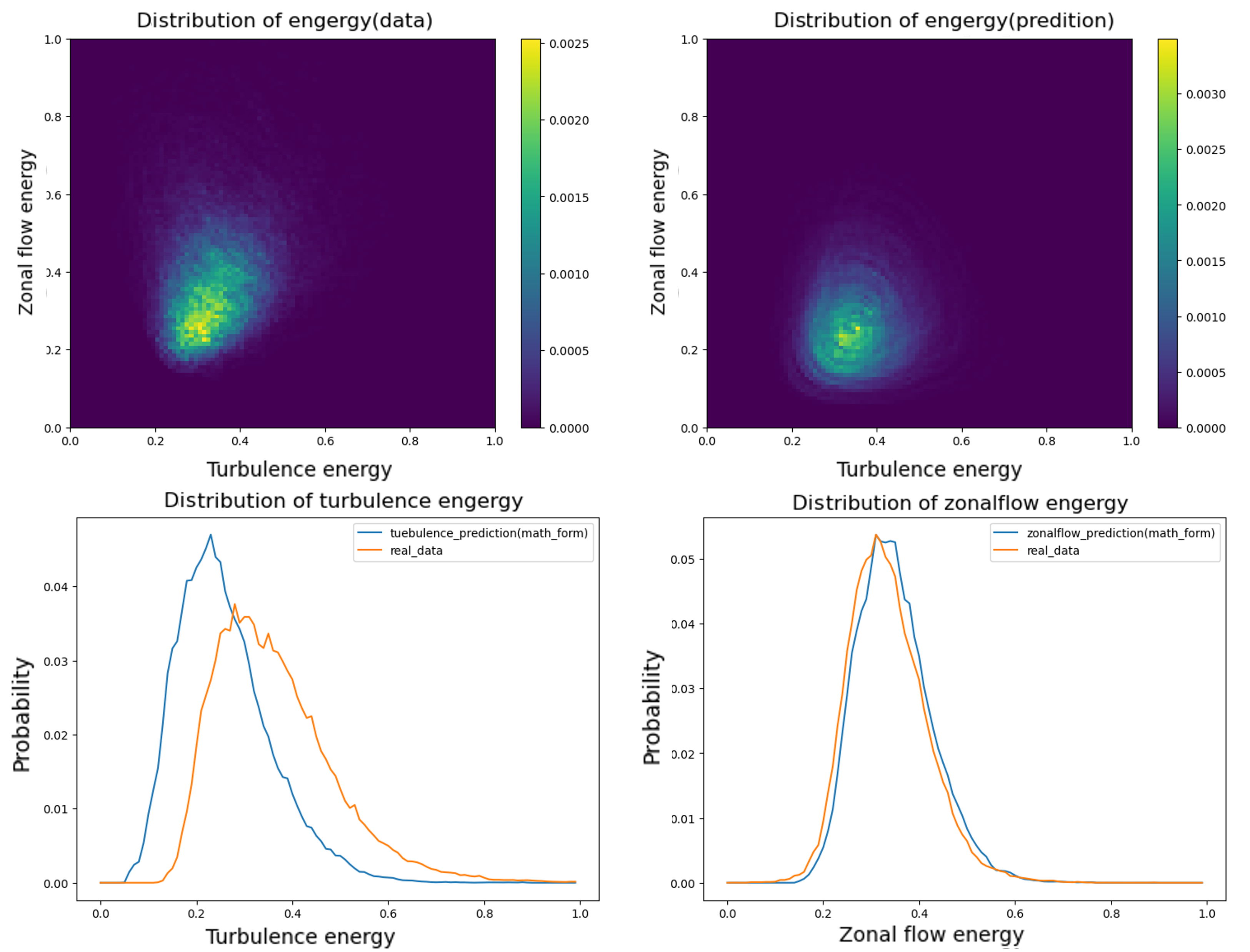} 
            \caption{Predicted states density distribution from mathematical form}
            \label{madis_old.fig}
        \end{figure}

        \indent We also do simulations with mathematical form \eqref{mathform}, and the density distribution of 50 simulations of 10,000 time steps in mathematical form \eqref{mathform} shows a quite similar shape in state distribution but with a shifted center. The former problem of constrained states density in central area for NN model is fixed after converted to mathematical form, see \figref{madis_old.fig}, because linear regression extracted the main features of the model and make up sacrificed information mentioned above through mathematical form. However, another problem arises, its performance is worse compared to before, a means Kullback–Leibler (KL) divergences $(0.82\pm0.06)$ due to a shifted center.
        \\
        \indent In fact, additional errors are introduced in the process of converting it to mathematical form, and such error would play a quite significant role in stochastic simulation, as a small deviation would cause an apparent sift in distribution; see \figref{madis_old.fig}. As mentioned above, this stochastic systems can exhibit significant variations even with small differences (as little as $2\%$). To fix this shift center problem, NN need more details provided.\\
        \indent To refine our approach, we further introduce a neural network with a more detailed structure, building on the results of the previous model to make finer adjustments. The modified network’s structure is illustrated in \figref{NN1.fig}, while all other settings remain the same as in the previous neural network. More details are discussed in \ref{subr}.
\section{Results for parameter scan}
    \label{app:scan_results}
    \begin{flushleft}
          \begin{tabular}{|l|c|c|c|r|}
            \hline
            $\kappa$ & $f_{11}(U)$ & $f_{12}(U)$ & Coefficient $\mathbf{A}$ & Coefficient $\mathbf{B}$ \\
            \hline
            0.95 & $0.10 U^2 - 0.146U + 0.029$ & $-0.048U^2 - 0.0133U + 0.05$ & 0.003491 & 0.01462141\\
            \hline
            1.00 & $0.14U^2 - 0.164U + 0.035$ & $-0.148U^2 + 0.065U + 0.03$ & 0.00462468 & 0.01337141\\
            \hline
            1.05 & $0.13U^2 - 0.165U + 0.040$ & $-0.244U^2 + 0.15U + 0.016$ & 0.0020408 & 0.01422784\\
            \hline
            1.10 & $0.14U^2 - 0.18U + 0.0466$ & $-0.183U^2 + 0.127U + 0.010$ & 0.00807797 & 0.01212985\\
            \hline
            1.15 & $0.096U^2 - 0.15U + 0.044$ & $-0.034U^2 + 0.0027U + 0.035$ & 0.01036318 & 0.01212277\\
            \hline
          \end{tabular}
    \end{flushleft}
\end{document}